\documentclass[a4paper,fleqn,usenatbib]{mnras}

\usepackage{newtxtext,newtxmath}

\usepackage[T1]{fontenc}
\usepackage{ae,aecompl}

\usepackage{graphicx}	
\usepackage{amsmath}	
\newcommand{\nicer}{\textit{NICER}}
\newcommand{\nustar}{\textit{NuSTAR}}
\newcommand{\swift}{\textit{Swift}}
\newcommand{\astrosat}{\textit{AstroSat}}
\newcommand{\maxi}{\textit{MAXI}}
\newcommand{\eps}{erg s$^{-1}$}
\newcommand{\ecs}{erg cm$^{-2}$ s$^{-1}$}
\newcommand{\pcm}{cm$^{-2}$}

\newcommand{\phc}{ph cm$^{-2}$ s$^{-1}$}
\newcommand{\kms}{km s$^{-1}$}

\newcommand{\gpc}{g cm$^{-3}$}
\newcommand{\ecps}{erg cm s$^{-1}$~}
\newcommand{\source}{MAXI~J1803--298}

\title[MAXI~J1803--298]{AstroSat Observation of X-ray Dips and State Transition in the Black Hole Candidate MAXI~J1803--298}
\author[Jana et al.]{
Arghajit Jana$^{1,2}$\thanks{E-mail: argha0004@gmail.com},
Sachindra Naik$^{1}$,
Gaurava K. Jaisawal$^{3}$,
Birendra Chhotaray$^{1,4}$,
Neeraj Kumari$^{1,4}$,
\newauthor
Shivangi Gupta$^{5}$
\\
$^{1}$Astronomy \& Astrophysics Division, Physical Research Laboratory, Navrangpura, Ahmedabad, 380 009, India\\
$^{2}$Institute of Astronomy, National Tsing Hua University, Hsinchu 30013, Taiwan \\
$^{3}$National Space Institute, Technical University of Denmark, Elektrovej, 327-328, DK-2800 Lyngby, Denmark \\
$^{4}$Indian Institute of Technology, Gandhinagar, 382 355, India\\
$^{5}$UM-DAE Centre for Excellence in Basic Sciences, University of Mumbai, Vidyanagari Campus, Mumbai, 400098, India\\
}
\date{Accepted XXX. Received YYY; in original form ZZZ}

\pubyear{2020}

\begin{document}

\label{firstpage}
\pagerange{\pageref{firstpage}--\pageref{lastpage}}
\maketitle

\begin{abstract}
We present the results obtained from broadband X-ray timing and spectral analysis of black hole candidate MAXI~J1803--298 using an {\it AstroSat} observation on May 11--12, 2021. Four periodic absorption dips with a periodicity of $7.02 \pm 0.18$~hour are detected in the light curve. {\it AstroSat} observe the source when it was undergoing a transition from hard-intermediate state to soft-intermediate state. Our timing analysis reveals the presence of a sharp type-C quasi periodic oscillation (QPO) in the power density spectra (PDS) with an evolving QPO frequency ranging from $5.31 \pm 0.02$~Hz to $7.61\pm 0.09$~Hz. We investigate the energy dependence of the QPO and do not find this feature in the PDS above 30~keV. The combined $0.7-80$~keV SXT and LAXPC spectra are fitted with a model consisting of thermal multi-colour blackbody emission and Comptonized emission components. We perform time-resolved spectroscopy by extracting spectra during the dip and non-dip phases of the observation. A neutral absorber is detected during the dip and non-dip phases though a signature of an ionized absorber is also present in the dip phases. The spectral and temporal parameters are found to evolve during our observation. We estimate the mass function of the system as $f(M) = 2.1-7.2~M_{\odot}$ and the mass of the black hole candidate in the range of $M_{\rm BH} \sim 3.5-12.5~M_{\odot}$.
\end{abstract}


\section{Introduction}
\label{sec:intro}
An X-ray binary (XRB) system consists of a compact object which is either a black hole (BH) or a neutron star (NS) and a normal companion star. Depending on the mass of the companion star, an XRB can be classified as low mass X-ray binary (LMXB) or high mass X-ray binary (HMXB). Generally, an LMXB contains a A or later type star, while an HMXB contains a giant O or B type star  \citep{White1995,Tetarenko2016}. An XRB can also be termed as a transient or persistent source, depending on its X-ray activity. A transient source spends most of the time in quiescence phase with X-ray luminosity, $L_{\rm X} < 10^{32}$ \eps and show occasional outbursts when the X-ray intensity rises by about three orders of magnitude or more compared to the quiescence phase \citep{RM06}. Instead, a persistent source is always active in X-rays with X-ray luminosity, $L_{\rm X} > 10^{35}$ \eps~ \citep{RM06}.

A black hole X-ray binary (BHXB) shows rapid variability in spectral and temporal properties \citep{Mendez1997,vanderklis1994}. The spectrum of a BHXB can be approximated by a multi-colour blackbody and a power-law tail. The multi-colour blackbody component is understood to be originated in a standard geometrically thin and optically thick disc \citep{SS73} while the power-law tail is believed to originate in a hot electron cloud known as Compton cloud or corona \citep{HM93,T94,CT95,Done2007}. The soft thermal photons originated in the disc undergo inverse-Compton scattering in the Compton cloud and produce hard Comptonized power-law tail \citep{ST80,ST85}. 

In BHXBs, the fast-variability in X-ray emission is generally seen in the power density spectra (PDS) of the source. The PDS are characterized with the so-called band-limited noise with a flat profile in the $\nu-p_{\nu}$ spectrum with a break frequency \citep{vanderklis1994}. The noise profile of PDS can be described by one or multiple Lorentzian functions \citep{Nowak2000,Belloni2002}. A BHXB also exhibits peaked noise or quasi-periodic oscillation (QPO) in the PDS. The low frequency QPO is generally observed in the range of $0.1-30$~Hz. Depending on the Q-factor (Q=$\nu /\Delta \nu$, $\nu$ and $\Delta \nu$ are centroid frequency and full-width at half maximum; FWHM), the rms amplitude of the QPO and broadband noise, a low frequency QPO (LFQPO) can be classified as type-A, type-B or type-C \citep[][and references therein]{Casella2005}. 

A correlation between the spectral and timing properties of a BHXB is observed in the hardness-intensity diagram \citep[HID;][]{Homan2001,Homan2005}, accretion rate ratio-intensity diagram \citep[ARRID;][]{Mondal2014,AJ2016}, rms-intensity diagram \citep[RID;][]{Munoz-Darias2011} or hardness ratio-rms diagram \citep[HRD;][]{Belloni2005}. An outbursting BHXB also shows different spectral states during the outburst, which is evident from the different branches of the HID, ARRID, RID or HRD. Generally, an outbursting BHXB evolves through low-hard state (LHS), hard-intermediate state (HIMS), soft-intermediate state (SIMS) and high soft state (HSS) during the outburst \citep{RM06,Nandi2012}. A detail study on the spectral state evolution has been made by several group for different outbursting BHXBs \citep{RM06,Tetarenko2016,Alabarta2020,KC2020}.

Recently, the \maxi/GSC discovered the new X-ray transient MAXI~J1803--298 when the source showed an X-ray outburst on May 1, 2021 \citep{Serino2021}. The source was subsequently observed with several other X-ray missions, e.g., \swift~ \citep{Gropp2021}, \nicer~ \citep{Gendreau2021,Bult2021}, \astrosat~ \citep{AJ2021b}, \nustar~ \citep{Xu2021}, {\it INTEGRAL} \citep{Chenevez2021} and {\it Insight}-HXMT \citep{Wang2021}. Following the discovery, optical \citep{Buckley2021,Hosokawa2021} and radio \citep{Espinasse2021} observations of the source were also carried out during the X-ray outburst. \nicer, \nustar~ and \astrosat~ observations revealed the presence of periodic absorption dips with a periodicity of $\sim 7$~hrs in the light curves of \source~ \citep{Homan2021,Xu2021,AJ2021b}. QPOs were also detected in the PDS of \source~ \citep{Chand2021,Ubach2021,Wang2021}. During \maxi/GSC observations, state transition of the source from the LHS to HSS was observed \citep{Shidatsu2021}. The \swift/XRT observations showed the evidence of disc wind in the source \citep{Miller2021}. A preliminary spectral analysis suggested that the source is a black hole X-ray binary \citep{Xu2021,Shidatsu2021,AJ2021b}.

In this paper, we present the results obtained from the \astrosat~ target of opportunity (TOO) observation of  \source~ during the recent X-ray outburst. Here, we study the spectral and timing properties of the source in detail to understand the accretion dynamics. Along with the \astrosat~ observation, we also used \nustar~ observation for the study of periodicity and dips in the source light curve. The paper is organized in the following way. In \S2, we present the observation, data reduction and data analysis process. In \S3, we present the results from a detailed spectral and timing studies of the source. We discuss our findings in \S4 and summarized the work in \S5.

\section{Observations and Data Analysis}
\label{sec:obs}
Following the discovery, \source\ was observed with \astrosat~ on two epochs i.e., on  May $11-12$, 2021 and July $24-26$, 2021 for exposures of $\sim 50$~ks each. \nustar~ observed the source on May 5, 2021 for an exposure of $\sim 27$~ks. The log of the observation is presented in Table~\ref{tab:log}.

\begin{table}
\caption{Observation log}
\centering
\begin{tabular}{lcccc}
\hline
Observation ID & Start Date & End Date & Exposure (ks) \\
 & Start time & End time & LAXPC/SXT \\
\hline
T04$\_$006T01$\_$9000004370  & 2021-05-11 & 2021-05-12 & 50 / 27.8 \\
  & 12:05:30 & 13:03:01      & \\
T04$\_$028T01$\_$9000004584 & 2021-07-24 &  2021-07-26 & 50 / 27.8\\
 & 15:48:09 & 00:59:02 &   \\
\hline
90702316002$^*$ & 2021-05-05 & 2021-05-06 & 27$^*$ \\
 & 16:46:09 & 10:26:09 & \\
\hline
\end{tabular}
\leftline{$^*$ \nustar~ observation.}
\label{tab:log}
\end{table}

\subsection{AstroSat}
\label{sec:data-red}
\astrosat~ is the first Indian multi-wavelength astronomical satellite launched by Indian Space Research Organization on September 28, 2015 \citep{Agrawal2006}. It provides a broad-band coverage from optical to hard X-ray bands for exploring the nature of the cosmic sources. There are five sets of instruments such as Soft X-ray Telescope \citep[SXT;][]{Singh2017}, Large Area X-ray Proportional Counters \citep[LAXPC;][]{Agrawal2017,Antia2017}, Cadmium Zinc Telluride Imager \citep[CZTI;][]{Rao2017}, a Scanning Sky Monitor \citep[SSM;][]{Ramadevi2018}, and Ultraviolet Imaging Telescope \citep[UVIT;][]{Tandon2017}, onboard the satellite. In the present work, we did not use data from the CZTI as the source was relatively faint for the detector. 

The SXT is a soft X-ray focusing telescope consisting of a CCD detector. It is sensitive in $0.3-8$~keV energy range. The effective area and energy resolution of the instrument are 128 cm$^{-2}$ and 5–6\% at 1.5~keV and 22 cm$^{-2}$ and 2.5\% at 6 keV, respectively. \source~ was observed with SXT in the photon counting (PC) mode at a time resolution of 2.4~s. The level-1 data were processed using standard SXT pipeline software {\tt AS1SXTLevel2}-1.4b \footnote{\url{https://www.tifr.res.in/~astrosat_sxt/sxtpipeline.html}} to obtain cleaned event files from each orbit of observation. Thereafter, each orbit data are merged into a single cleaned event file using the {\tt sxtevtmergertool}  \footnote{\url{http://astrosat-ssc.iucaa.in/sxtData}} task. For time-resolved spectroscopy, we divided the event files into 10 segments using the {\tt XSELECT} package. As the source was bright in soft X-rays (count rate of $>40$ count/sec), the SXT data are affected by photon pile-up\footnote{\url{https://www.tifr.res.in/~astrosat_sxt/instrument.html}}. We chose an annular region with a fixed outer radius of 10 arcmin and a variable inner radius. The pile-up was checked from the spectral distortion\footnote{\url{https://www.swift.ac.uk/analysis/xrt/pileup.php}}. We found that the pile-up was removed when the inner radius of 8 arcmin is considered in our analysis. We used background spectra and response matrix files (RMF) that were supplied by the SXT instrument team \footnote{\url{https://www.tifr.res.in/~astrosat_sxt/dataanalysis.html}}. The auxiliary response file (ARF) is generated by the \textsc{sxtARFModule} tools.

\astrosat~ has three LAXPC units that are sensitive to the X-ray photons in $3-80$~keV energy range, with a total effective area of 8000 cm$^{-2}$ at 15~keV. The timing and spectral resolutions of the LAXPC are 10$\mu$S and 12\% at 22~keV, respectively. We used only event mode data from LAXPC20 unit in our analysis. We did not use data from LAXPC10 and LAXPC30 due to high background and gain issues with the instruments \citep{Antia2017,Antia2021}. We used standard data analysis tools {\tt LAXPCsoftware}\footnote{\url{http://astrosat-ssc.iucaa.in/laxpcData}} (\textsc{laxpcsoft}; version 2020 August 4) to extract light curves and spectra. The background estimation was done based on the blank sky observations, closest to the time of observation of the source of interest, as described in \citet{Antia2017}. In order to construct the response matrix, three on-board radioactive sources, covering different energies in the range of LAXPC detectors, were used \citep[see][for details]{Antia2017}. The spectral responses were generated carefully by appropriately modeling and accounting for known 30 keV fluorescence photons produced due to Xe-K shell interaction of incident X-rays. We applied barycentric correction on the background subtracted light curves using {\tt as1bary} tool. The LAXPC light curve is also divided into 10 segments for studying the source properties during dip and non-dip phases. The energy resolved light curves were generated using the tools provided by {\tt LAXPCsoftware}. The second \astrosat~ observation was carried out when the source was in quiescent phase. Therefore, the LAXPC data are background dominated above $\sim 10$~keV. Though we generated light curves and spectra from this observation, it was difficult to perform a detailed timing and spectral analysis of the data due to low source flux.

\subsection{NuSTAR}
\label{sec:nustar-red}
\nustar~ is a hard X-ray focusing telescope, consisting of two identical modules: FPMA and FPMB \citep{Harrison2013}. \nustar~ observed \source~ on May 5, 2021 for a exposure of $\sim 27$~ks (see Table~\ref{tab:log}). We reprocessed the raw data with the \nustar~ Data Analysis Software ({\tt NuSTARDAS}, version 1.4.1). Cleaned event files were generated and calibrated by using the standard filtering criteria in the {\tt nupipeline} task and the latest calibration data files available in the NuSTAR calibration database (CALDB) \footnote{\url{http://heasarc.gsfc.nasa.gov/FTP/caldb/data/nustar/fpm/}}. The source and background products were extracted by considering circular regions with radii 60 arcsec and 90 arcsec, respectively. The light curves were extracted using the {\tt nuproduct} task. For the present work, we only used \nustar~ light curve to study the periodicty and dips.

\section{Analysis and Results}
\label{sec:res}

\subsection{Outburst Profile}
\label{sec:prof}

\begin{figure}
\centering
\includegraphics[width=8.5cm]{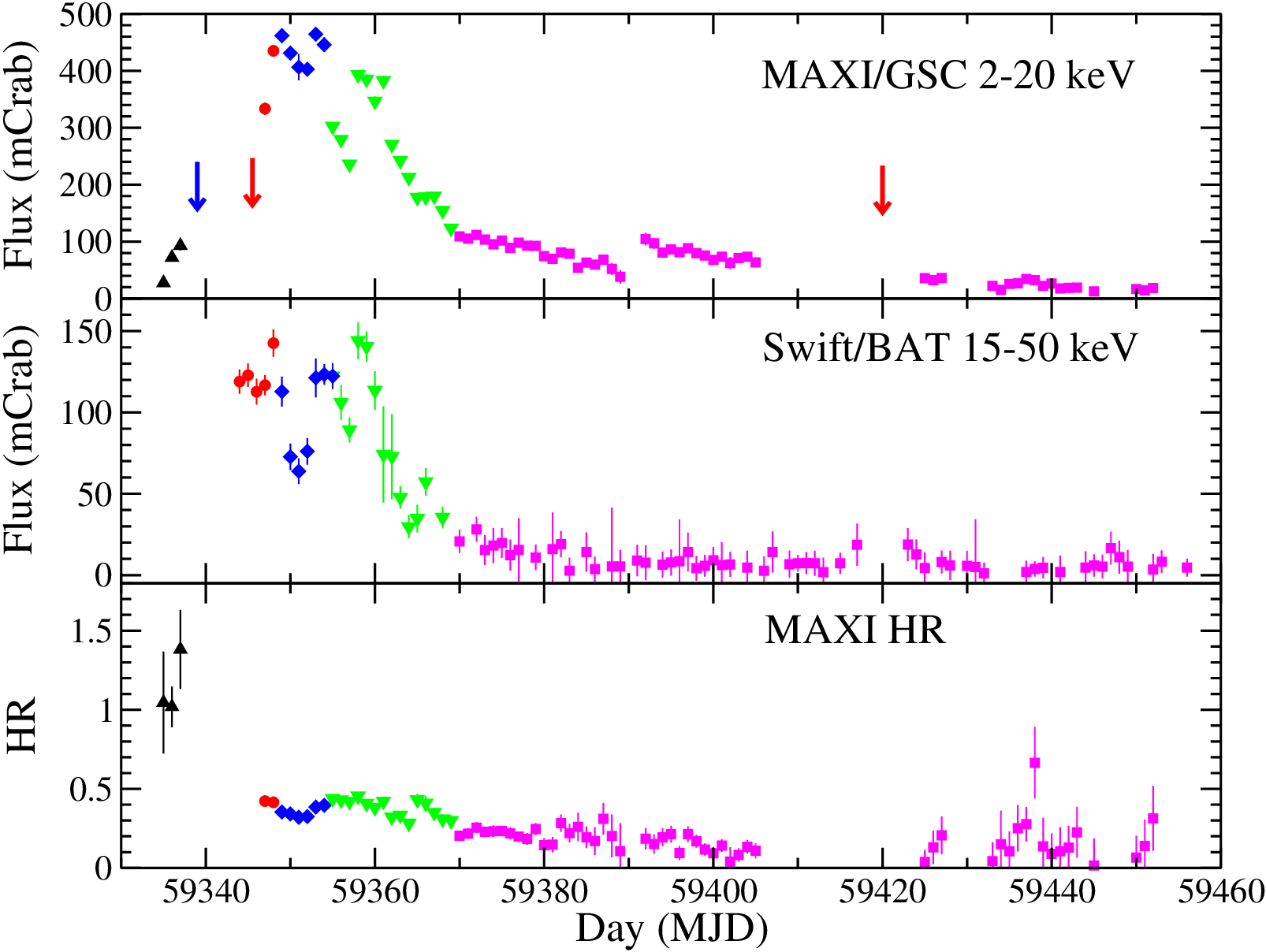}
\caption{Top : The $2-20$~keV \maxi/GSC light curve of \source~ covering the recent X-ray outburst. Middle: The $15-50$~keV \swift/BAT light curve for the duration shown in the top panel. Bottom: The hardness ratio (HR1 = ratio between the count rates in 4--10~keV and 2--4 keV ranges). The black triangles, red circles, blue diamonds, green down-triangles and purple squares represent the data points from the LHS (rising), IMS (ris.), HSS, IMS (decl.) and LHS (decl.), respectively. The red and blue arrows mark the \astrosat~ and \nustar~ observations of the source, respectively.}
\label{fig:profile}
\end{figure}

\source~ was observed extensively with several X-ray missions following its discovery on May 1, 2021. The source was in the outbursting phase for about three months. The outburst profile of the source is shown in Figure~\ref{fig:profile}. Top and middle panels of Figure~\ref{fig:profile} show the $2-10$~keV \maxi/GSC and $15-50$~keV \swift/BAT light curves, respectively. The bottom panel of Figure~\ref{fig:profile} shows the variation of the hardness ratio (HR1) of the source during the outburst. The HR1 is defined as the ratio between the count rates in $4-10$~keV and $2-4$~keV ranges of 
\maxi/GSC. Symbols with different colours represent different spectral states of the source.

In the beginning of the outburst, the source intensity increased rapidly. The intensity in $2-10$~keV range reached its peak on May 15, 2021 (MJD 59349) and then decreased until May 18, 2021 (MJD 59352). The $15-50$~keV BAT flux reached its peak ($\sim 142$ mCrab) on May 14, 2021 (MJD 59348) and then decreased to a local minimum value of $\sim 63$ mCrab on May 17, 2021 (MJD 59351). However, the $2-20$~keV X-ray light curve showed another peak on May 19, 2021 (MJD 59353). After that, the $2-20$~keV flux decreased to a minimum on May 23, 2021 (MJD 59357), followed by a temporary increase next day. The source intensity then monotonically decreased until June 3, 2021 (MJD 59368). The $15-50$~keV BAT flux also showed similar variation. Beyond June 4, 2021 (MJD 59369), both $2-20$~keV GSC flux and $15-50$~keV BAT flux decreased slowly. A minor flare like event was observed around June 27, 2021 (MJD 59392). After that, the X-ray intensity decreased gradually and the source moved to the quiescent state.

Based on the variation of X-ray intensity and HR1 during the outburst, it appears that the source transited through three canonical spectral states. The source evolved through the LHS (rising) $\rightarrow$ HIMS (ris.) $\rightarrow$ IMS (ris.) $\rightarrow$ HSS $\rightarrow$ IMS (decl.) $\rightarrow$ LHS (decl.). In the beginning of the outburst, the source was in the LHS (ris.). It moved into the HSS through the IMS (ris.) and then entered the HSS on May 15, 2021 (MJD 59349) when the $2-20$~keV flux was maximum. The source, then, moved into the IMS (decl.) in the decay phase when the X-ray intensity rapidly decreased until June 3, 2021 (MJD 59368). After this, the source was in the LHS (decl.) before moving to the quiescent state.

\begin{figure*}
\centering
\includegraphics[width=18cm]{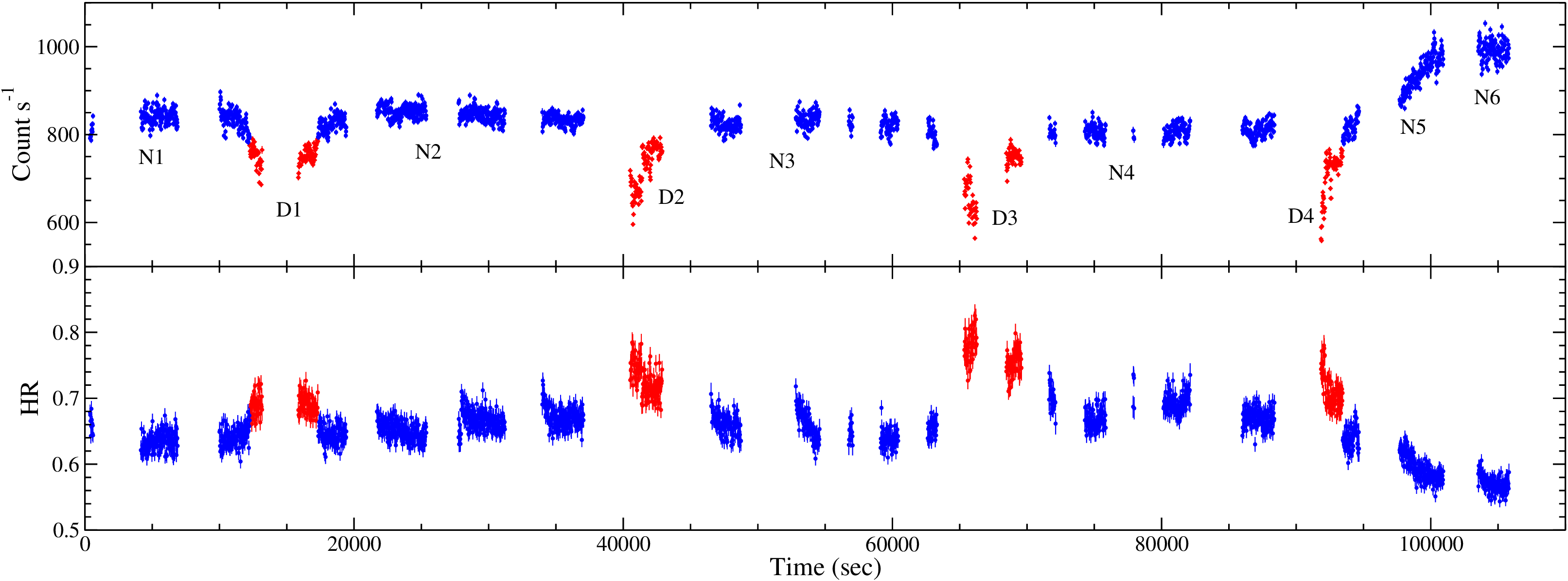}
\caption{Top : The $3-80$~keV LAXPC light curve of the source from the \astrosat~ observation on May 11, 2021. Bottom : The variation of hardness ratio (HR2 = $6-30$~keV counts / $3-6$~keV count rates). The blue and red points indicate the data from the non-dip and dip phases, respectively.}
\label{fig:lxp-lc}
\end{figure*}

\begin{figure}
\centering
\includegraphics[width=8.5cm]{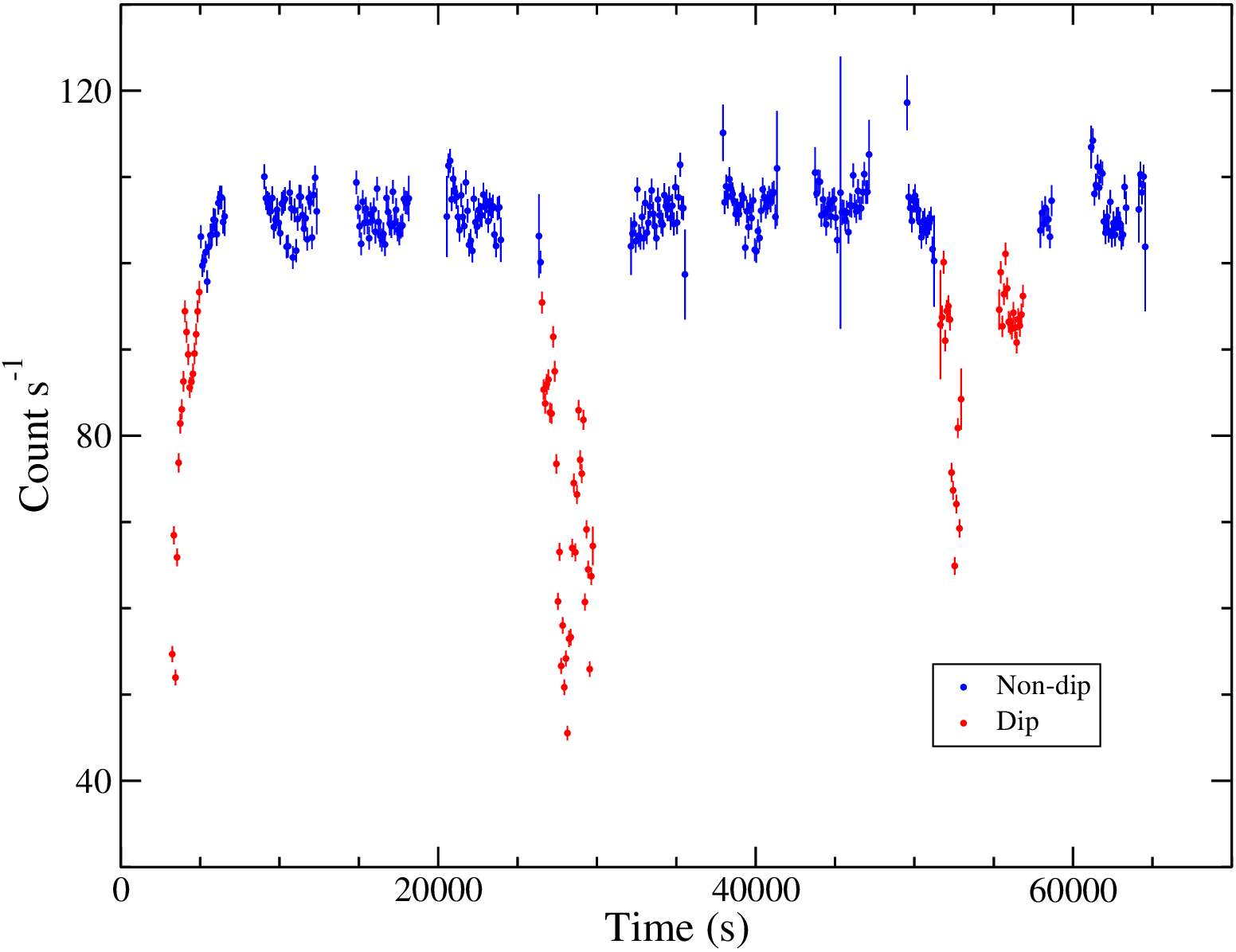}
\caption{The $3-78$~keV light curve of \source~ from the \nustar~ observation on May 5, 2021. The blue and red points indicate data from the non-dip and dip phases, respectively.}
\label{fig:nust-lc}
\end{figure}

\begin{figure}
\centering
\includegraphics[width=8.5cm]{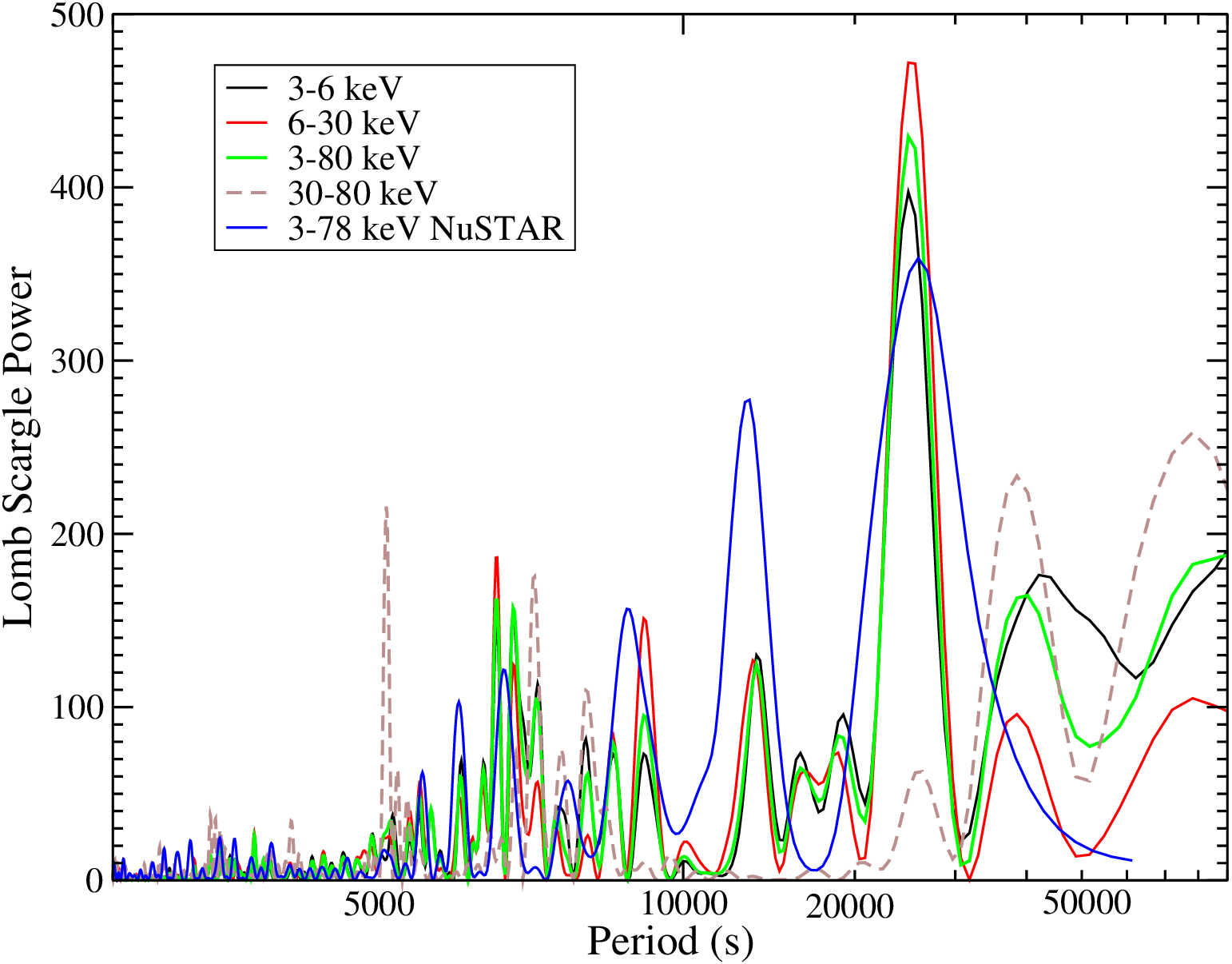}
\caption{Lomb-Scargle (LS) periodogram analysis. The black, red, green and grey lines represent the LS periodograms obtained from the light curves in $3-6$~keV, $6-30$~keV, $3-80$~keV and $30-80$~keV ranges, respectively. The blue solid line represents the $3-78$~keV \nustar~ lightcurve. A clear periodicity at $\sim$25200 s is detected in all light curves except for the $30-80$~keV LAXPC light curve.}
\label{fig:LS}
\end{figure}

\subsection{Absorption Dips and Periodicity}
\label{sec:abs-dips}
\astrosat~ observed \source~ on May 11--12, 2021 and July 24--26, 2021 for a net exposure of 100~ks. The red arrows in the top panel of Figure~\ref{fig:profile} mark the \astrosat~ observations. The first observation was carried out when the source was in the rising phase of its outburst, while the second observation was made when the source was in the quiescent state. The background subtracted $3-80$ keV light curve of the source from the first observation is shown in the top panel of Figure~\ref{fig:lxp-lc}. The variation of HR2 ($6-30$~keV/$3-6$~keV count rates) is shown in the bottom panel of Figure~\ref{fig:lxp-lc}. Four absorption dips, marked as D1, D2, D3 \& D4, are clearly visible in the LAXPC light curve. The dip regions are marked with red circles while the blue circles mark the non-dip regions. Beside the four dips, we divided the non-dips regions into six segments (marked as N1, N2, N3, N4, N5 \& N6). The last two segments (N5 \& N6) were treated separately as N5 \& N6 showed different trend of X-ray intensity. The X-ray intensity increased during the N5, while it was constant during N6. Higher values of hardness ratio (HR2) are seen during the dip regions (see bottom panel of Figure~\ref{fig:lxp-lc}). This indicates that the soft X-ray photons in $3-6$~keV range were primarily affected at the dip phases in comparison to the hard X-ray photons in $6-30$~keV range. Figure~\ref{fig:nust-lc} shows the $3-78$~keV light curve obtained from the \nustar~ observation. The dips are also visible in the \nustar~ lightcurve.

The dips in the light curve (top panel of Figure~\ref{fig:lxp-lc} and Figure~\ref{fig:nust-lc}) appear to be periodic. To determine the periodicity, we applied Lomb-Scargle (LS) periodogram method to calculate the periodicity in the light curves at different energy bands, e.g. $3-80$~keV, $3-6$~keV, $6-30$~keV, $30-80$~keV, $3-78$~keV (for \nustar) ranges \citep{Lomb1976,Scargle1982,VanderPlas2018} using the \textsc{scargle} routine of the Starlink software\footnote{\url{http://starlink.jach.hawaii.edu/starlink}}. We searched all the lightcurves for periodic signals between 40s and 100000s. Figure~\ref{fig:LS} shows the LS periodograms of light curves in different energy bands. The black, red and green solid lines represent the periodograms of $3-6$~keV, $6-30$ and $3-80$~keV LAXPC light curves, respectively. The grey dashed line represents the periodogram obtained from the $30-80$~keV range light curve. The blue line represents the periodogram of \nustar~ observation in $3-78$~keV energy range.

We obtained the periodicity as $25187 \pm 711$~s ($7.00\pm 0.20$~hr), $25499 \pm 1333$~s ($7.08\pm 0.37$~hr) and $25267\pm 635$~s ($7.02\pm 0.18$~hr) in $3-6$~keV, $6-30$ and $3-80$~keV LAXPC light curves, respectively. The $3-78$~keV \nustar~ light curve shows the presence of a periodicity of $25830 \pm 3626$~s ($7.18\pm 0.92$~hr). Different energy band indicates a different periodicity. However, all the periodicities are consistent within the uncertainties. The discrepancy aroused due to the data gap in the light curve. The data gap due to the orbital motion of the satellite also did not allow us to estimate the exact duration of the dips. From visual inspection, the duration of the dips is $\sim 5000$~sec. We did not obtain any clear periodicity from the $30-80$~keV LAXPC light curve, indicating the photons above 30~keV were not affected during the dip phases. The detailed result of periodogram analysis is given in Table~\ref{tab:LS}.  We also calculated the false alarm probability (FAP) of the main peak in the LS periodogram considering 1000 trials \citep{VanderPlas2018} using \textsc{astropy} package\footnote{\url{https://docs.astropy.org/en/stable/timeseries/lombscargle.html.}} The FAP value is estimated to be 0.00-0.01 with 95\% confidence. The observed periodicity at 25187 s, 25499 s, 25267 s and 25830 s and were detected with a significance level of 95\% in the light curves in $3-6$~keV, $6-30$, $3-80$~keV and $3-78$~keV (\nustar~ observation) ranges, respectively.

The second \astrosat~ observation was carried out when the source was in the LHS/quiescent state with $2-20$ keV flux $\lesssim 30$~mCrab, about one order lower than the first \astrosat~ observation. The LAXPC data during this observation were background dominated above $\sim10$~keV. As the source was at a low flux level and LAXPC data were background dominated, we did not detect any dip in the light curve even in soft X-rays.

\begin{table}
\centering
\caption{Result of Lomb-Scargle Periodogram Analysis}
\begin{tabular}{lcccc}
\hline
Instrument & Energy band & Periodicity & LS power  \\
 & (keV) & (s) & \\
\hline
\astrosat/LAXPC & $3-6$ & $25187 \pm 711$ & 397 \\
\astrosat/LAXPC & $6-30$ & $25499 \pm 1333$ & 472 \\
\astrosat/LAXPC & $30-80$ & -- & -- \\
\astrosat/LAXPC & $3-80$ &  $25267\pm 635$ & 429\\
\nustar & $3-78$& $25830 \pm 3626$ & 359\\
\hline
\end{tabular}
\leftline{Errors are quoted at 1$\sigma$.}
\label{tab:LS}
\end{table}

\begin{figure*}
\centering
\includegraphics[width=18cm]{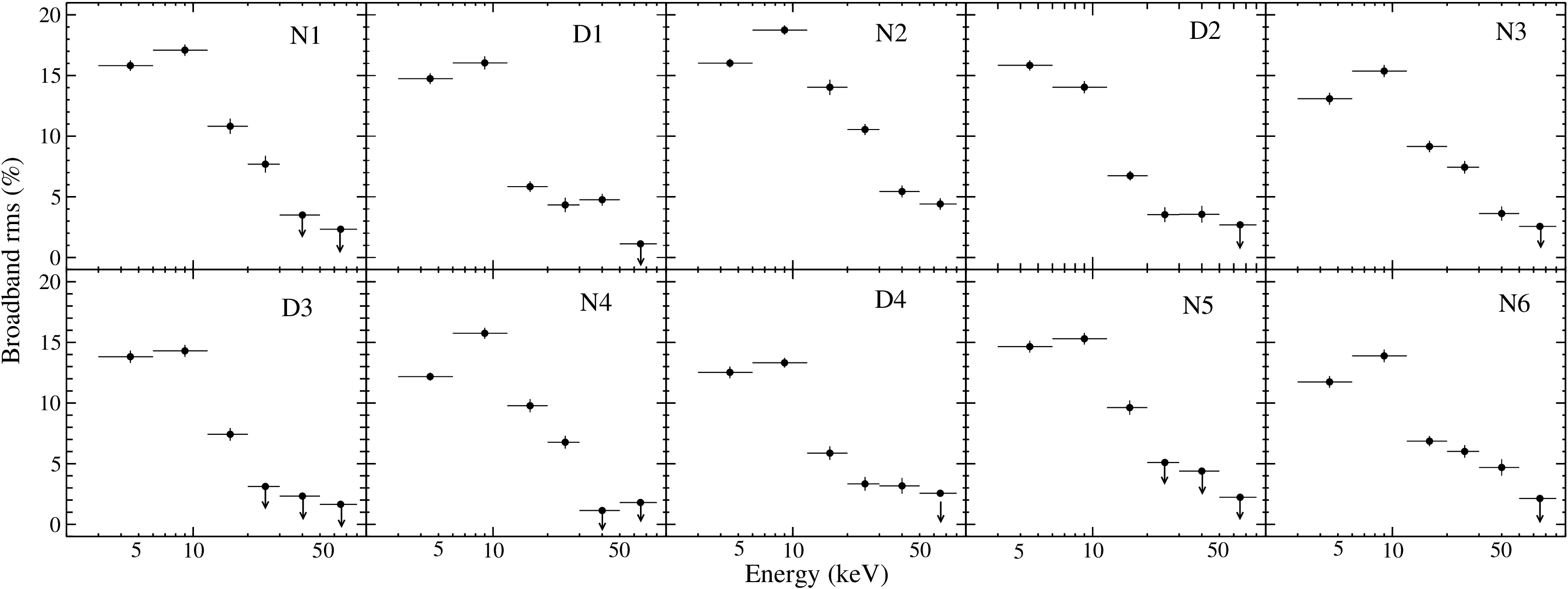}
\caption{Broadband RMS Spectra of \source~ during non-dip and dip regions of the light curve. The data points with down arrows indicate detection of upper limit.}
\label{fig:rms-spec}
\end{figure*}

\begin{figure*}
\centering
\includegraphics[width=16cm]{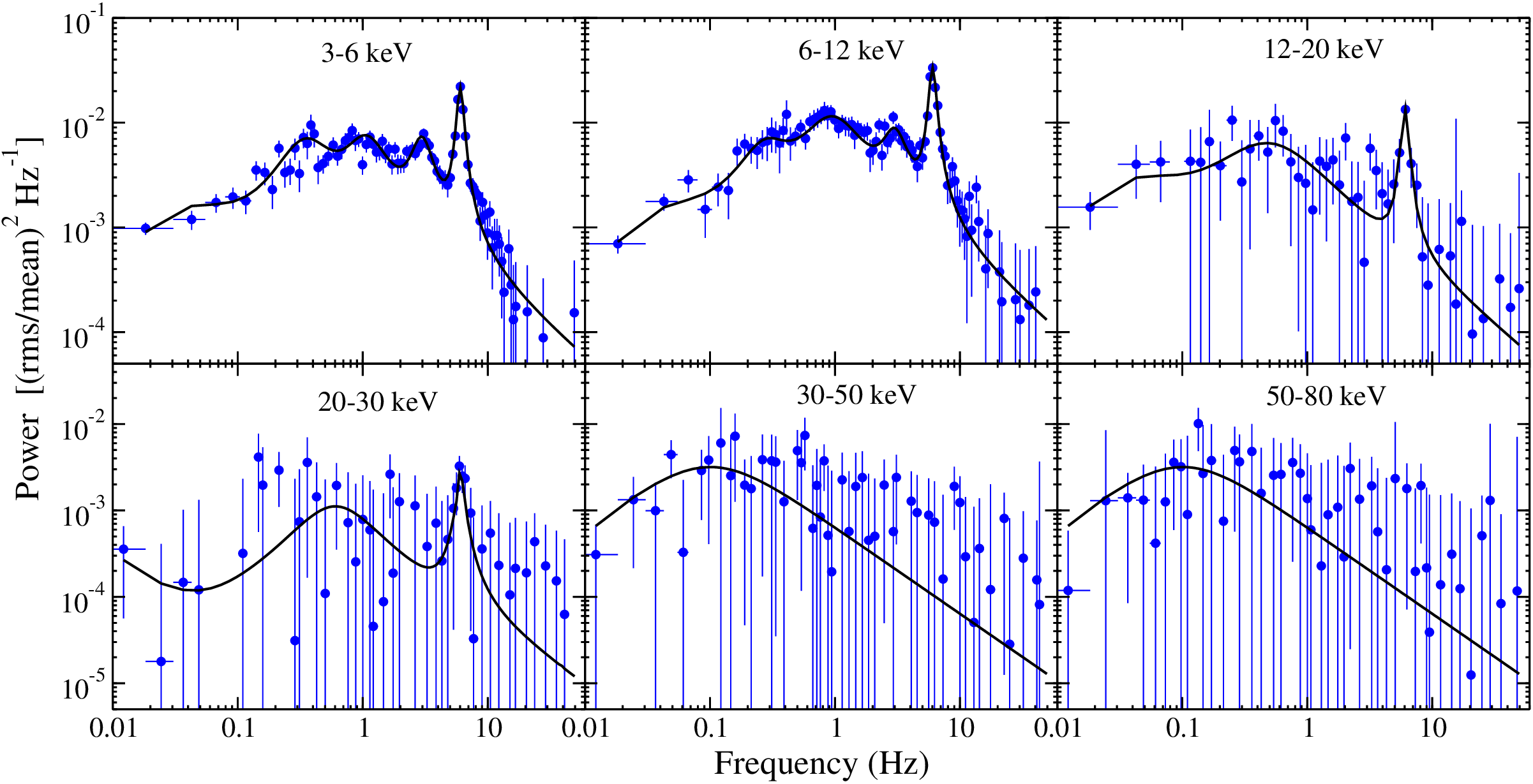}
\caption{Representative PDS for non-dip (N1) region in different energy bands. The black solid lines represent the best-fit Lorentzian function.}
\label{fig:pds-n1}
\end{figure*}

\subsection{Power Density Spectra}
\label{sec:pds}
We divided the entire LAXPC light curve from first \astrosat~ observation into ten segments. For each segment, we generated light curves at a binning time of 0.01~s in $3-80$~keV, $3-6$~keV, $6-12$~keV, $12-20$~keV, $20-30$~keV, $30-50$~keV and $50-80$~keV ranges. The power density spectra (PDS) are generated by applying the Fast Fourier Transformation (FFT) technique on the background subtracted LAXPC light curves using {\tt powspec} task of {\tt FTOOLS} for each segment. The light curves are divided into 8192 intervals and the Poisson noise subtracted PDS are generated for each interval. The final PDS are generated by averaging all the PDS from each interval. We re-binned the PDS in a geometrical series by using the normalization factor of 1.05. The final PDS are normalized to give the rms spectra in the unit of $(rms/mean)^2 Hz^{-1}$ \citep{Miyamoto1991,vanderklis1997}. The 0.01~s light curves allowed us to search for QPO in the PDS up to the Nyquist frequency of 50~Hz. 

We fitted all the PDS with a single or multiple Lorentzian functions, depending on the nature of the broadband noise (BN) and QPO. We calculated the fractional rms amplitude for broadband noise and QPO for each PDS by integrating the power in the range of $0.01-50$~Hz \citep{vanderklis2004}. We repeated this exercise for the light curves in different energy ranges for each segment. Using these rms values for different energy bands, we constructed rms spectra for each segment and showed in  Figure~\ref{fig:rms-spec} for all segments. The rms was found to be maximum (peak) in $6-12$~keV energy bands and then decreased with increasing energy. The detailed results of the PDS analysis is tabulated in Table~\ref{tab:qpo}.

\subsection{QPO Frequency}
\label{sec:qpo}

\begin{figure*}
\centering
\includegraphics[width=16cm]{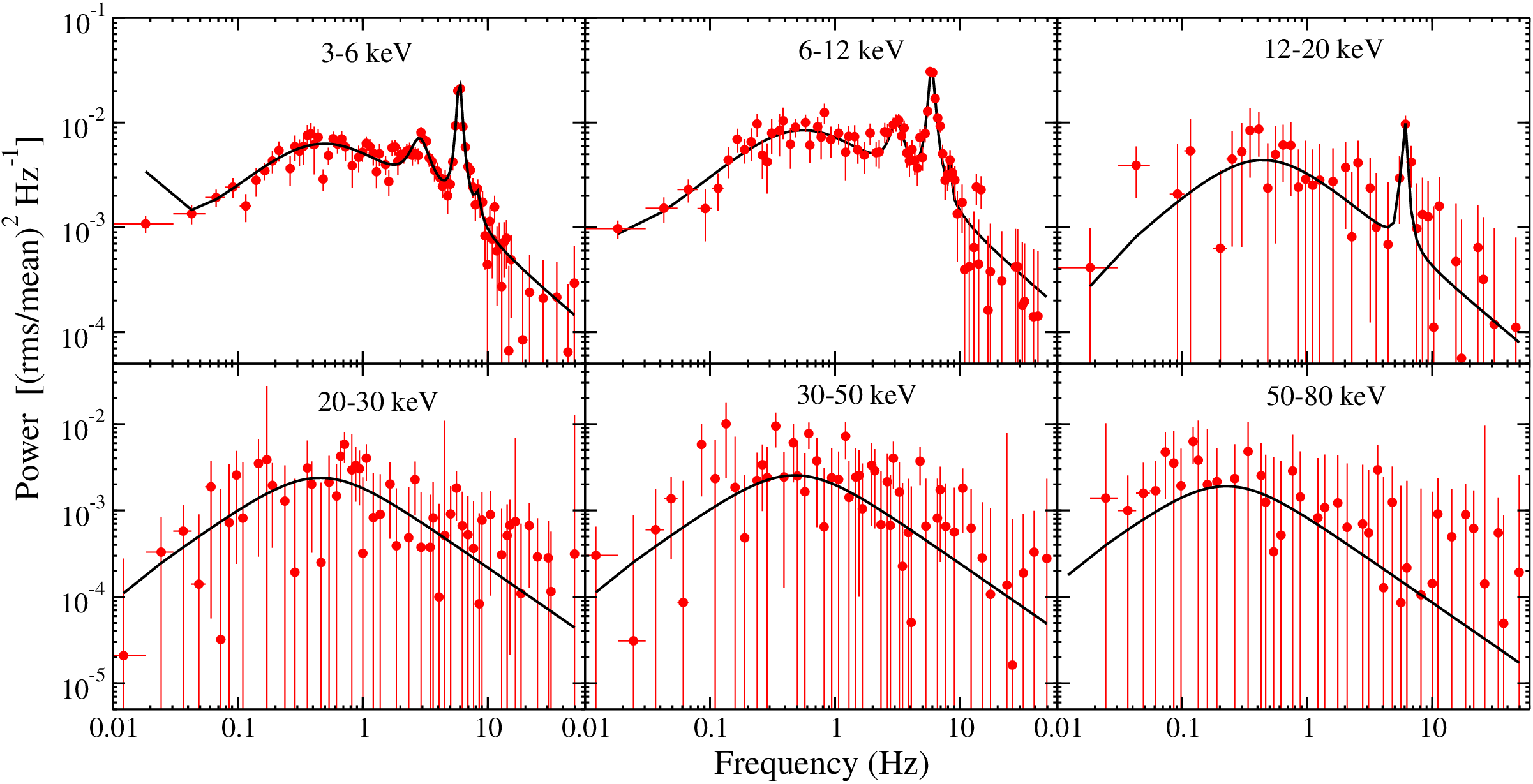}
\caption{Representative PDS for dip phase (D1) in different energy bands. The black solid lines represent the best-fit Lorentzian function.}
\label{fig:pds-d1}
\end{figure*}

\begin{figure}
\centering
\includegraphics[width=8.5cm]{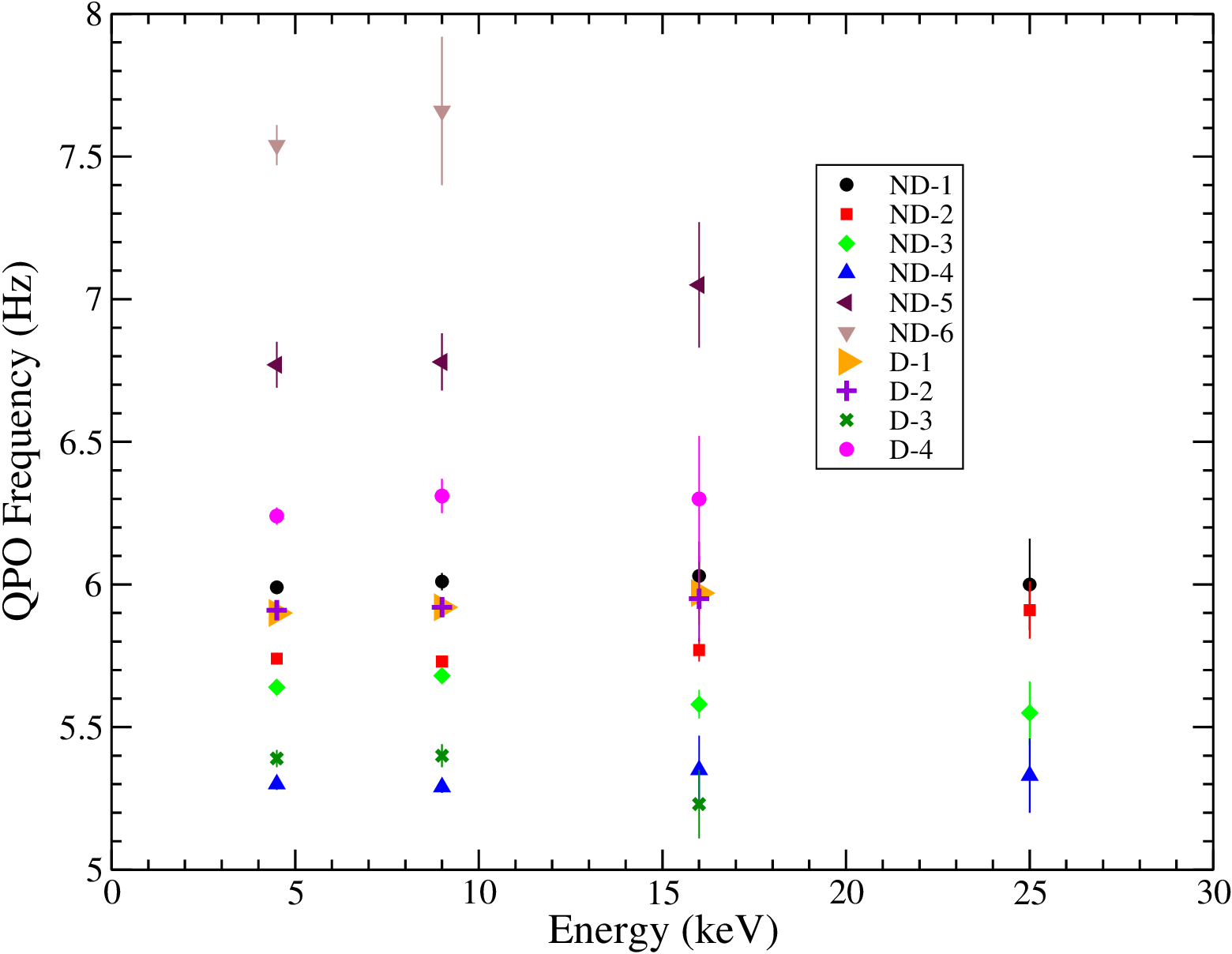}
\caption{Change is QPO frequency with energy. Different colour and symbols represent the observations from different segments.}
\label{fig:en-qpo}
\end{figure}

Figure~\ref{fig:pds-n1} and Figure~\ref{fig:pds-d1} show representative PDS in different energy bands for non-dip (N1) and dip (D1) regions, respectively. The PDS obtained from the $3-80$~keV light curves showed a sharp QPO at around $6$~Hz in every segment. On investigating the energy resolved PDS, we did not observe any variation of the QPO frequency with energy. The QPOs are observed in the energy ranges up to 30~keV for non-dip regions. However, in dip regions, QPOs were present in the PDS obtained from light curves up to 20~keV. Along with the primary QPO, a sub-harmonic was also observed in the PDS from light curves in $3-6$~keV and $6-12$~keV ranges. Figure~\ref{fig:en-qpo} shows the energy dependent QPO frequency for each segment. Table~\ref{tab:en-timing} presents the results from the energy-dependent timing analysis.

\begin{figure}
\centering
\includegraphics[width=8.5cm]{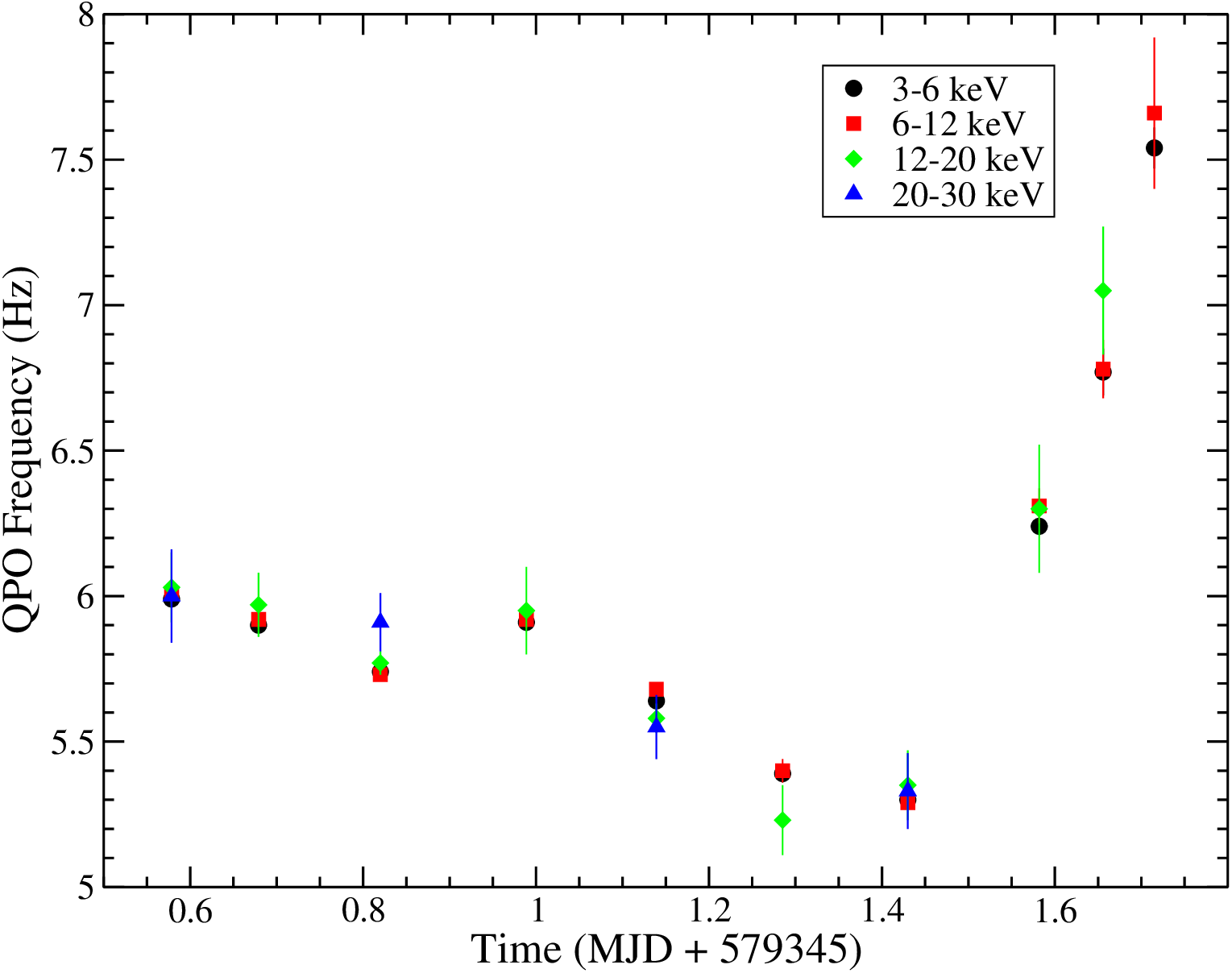}
\caption{Evolution of energy-dependent QPO with time during first AstroSat observation of \source. The black circles, red squares, green diamonds and blue triangles represent the QPO frequency in $3-6$~keV, $6-12$~keV, $12-20$~keV and $20-30$~keV energy bands, respectively.}
\label{fig:tim-qpo}
\end{figure}

\begin{table*}
\caption{Results obtained from the timing analysis of first \astrosat~ observation of \source.}
\centering
\begin{tabular}{lccccccccc}
\hline
Phase & Count Rate & QPO$^{\rm f}$ & Q-factor$^{\rm f}$ & rms$^{\rm f}$ & QPO$^{\rm h}$ & Q-factor$^{\rm h}$ & rms$^{\rm h}$ & BN rms \\
& (Count s$^{-1})$ & (Hz) & & (\%) & (Hz) & & (\%) & (\%) \\
\hline
N1&$728\pm 6$&$ 6.00\pm 0.02$&$  8.44\pm 0.07$&$ 7.21\pm 0.05$&$ 2.94\pm 0.05$&$ 2.19\pm 0.10$&$ 4.15\pm 0.09$&$ 12.27\pm 0.21$\\ 
D1&$666\pm 6$&$ 5.92\pm 0.02$&$ 10.08\pm 0.05$&$ 6.72\pm 0.05$&$ 2.96\pm 0.08$&$ 2.62\pm 0.12$&$ 3.33\pm 0.21$&$ 12.07\pm 0.24$\\
N2&$727\pm 6$&$ 5.75\pm 0.02$&$  7.05\pm 0.12$&$ 7.49\pm 0.07$&$ 2.81\pm 0.05$&$ 2.38\pm 0.13$&$ 2.79\pm 0.05$&$ 12.47\pm 0.19$\\
D2&$623\pm 6$&$ 5.94\pm 0.03$&$  9.60\pm 0.51$&$ 6.42\pm 0.04$&$ 2.89\pm 0.10$&$ 2.72\pm 0.14$&$ 3.13\pm 0.18$&$ 11.82\pm 0.18$\\
N3&$718\pm 6$&$ 5.66\pm 0.02$&$  6.20\pm 0.13$&$ 6.42\pm 0.12$&$ 2.74\pm 0.03$&$ 2.46\pm 0.11$&$ 2.93\pm 0.05$&$ 11.98\pm 0.15$\\
D3&$603\pm 6$&$ 5.39\pm 0.02$&$ 10.04\pm 0.06$&$ 6.34\pm 0.04$&$ 2.67\pm 0.16$&$ 3.10\pm 0.17$&$ 2.59\pm 0.37$&$ 11.71\pm 0.22$\\
N4&$699\pm 6$&$ 5.31\pm 0.02$&$  6.99\pm 0.12$&$ 5.90\pm 0.13$&$ 2.47\pm 0.04$&$ 2.21\pm 0.11$&$ 2.79\pm 0.07$&$ 11.79\pm 0.15$\\
D4&$793\pm 6$&$ 6.28\pm 0.03$&$  8.38\pm 0.08$&$ 4.92\pm 0.06$&$ 2.99\pm 0.12$&$ 2.65\pm 0.16$&$ 3.08\pm 0.18$&$ 10.85\pm 0.20$\\
N5&$876\pm 7$&$ 6.78\pm 0.08$&$  6.09\pm 0.21$&$ 6.13\pm 0.26$&$ 3.36\pm 0.23$&$ 2.02\pm 0.32$&$ 3.94\pm 0.07$&$ 9.46\pm 0.18$\\
N6&$992\pm 7$&$ 7.61\pm 0.09$&$  5.98\pm 0.22$&$ 2.86\pm 0.11$&$   -         $&$  -          $&$ -           $&$ 8.41\pm 0.18$\\
\hline
\end{tabular}
\leftline{The parameters corresponding to the fundamental component are indicated with superscript `f' and those corresponding to the sub-harmonic component}
\leftline{ are marked with `h'.}
\label{tab:qpo}
\end{table*}

We studied evolution of the low frequency QPO during our observation. We detected a $6.00\pm0.02$~Hz QPO in the PDS of $3-30$~keV range light curve from the non-dip region N1. The QPO frequency monotonically decreased to $5.31\pm 0.02$~Hz till non-dip region N4. After that, the QPO frequency increased rapidly, reaching $7.61\pm 0.96$~Hz during the region N6. Similar evolution of the QPO frequency were also seen in $3-6$ keV, $-12$ keV, $12-20$ keV and $20-30$ keV ranges. The evolution of the QPO frequency with time is shown in Figure~\ref{fig:tim-qpo}.

\begin{figure*}
\centering
\includegraphics[width=5.5cm]{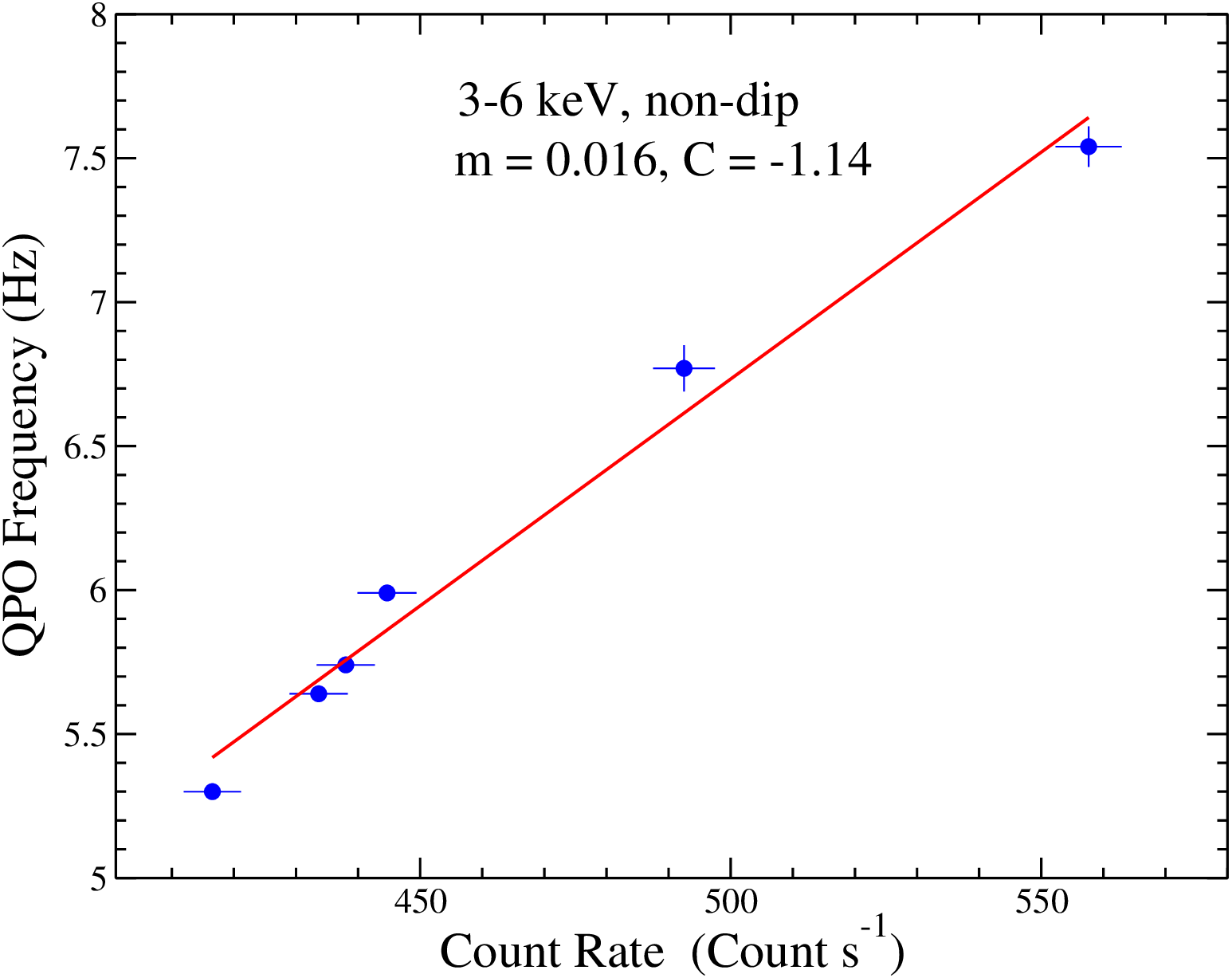}\hspace{0.2cm}
\includegraphics[width=5.5cm]{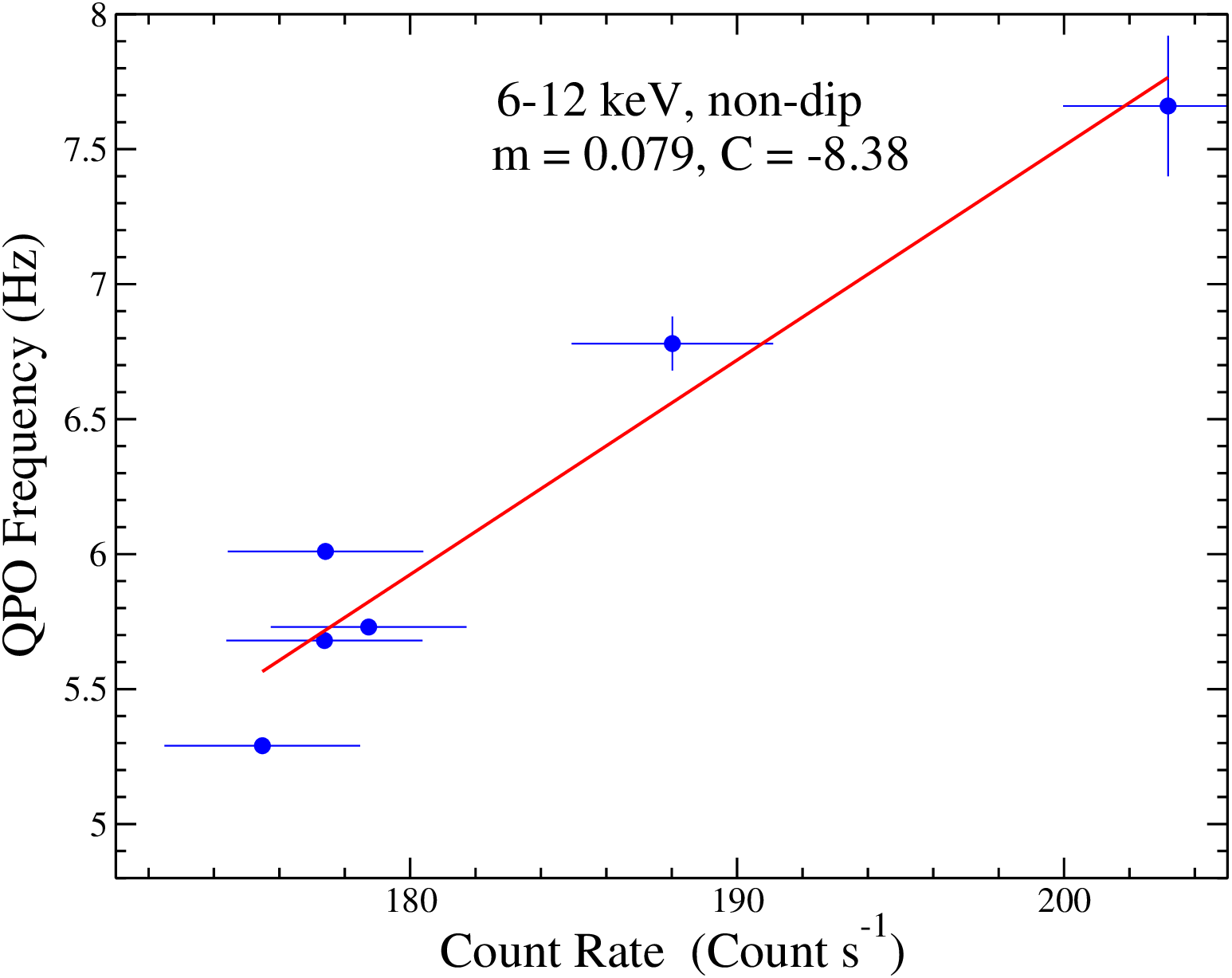}\hspace{0.2cm}
\includegraphics[width=5.5cm]{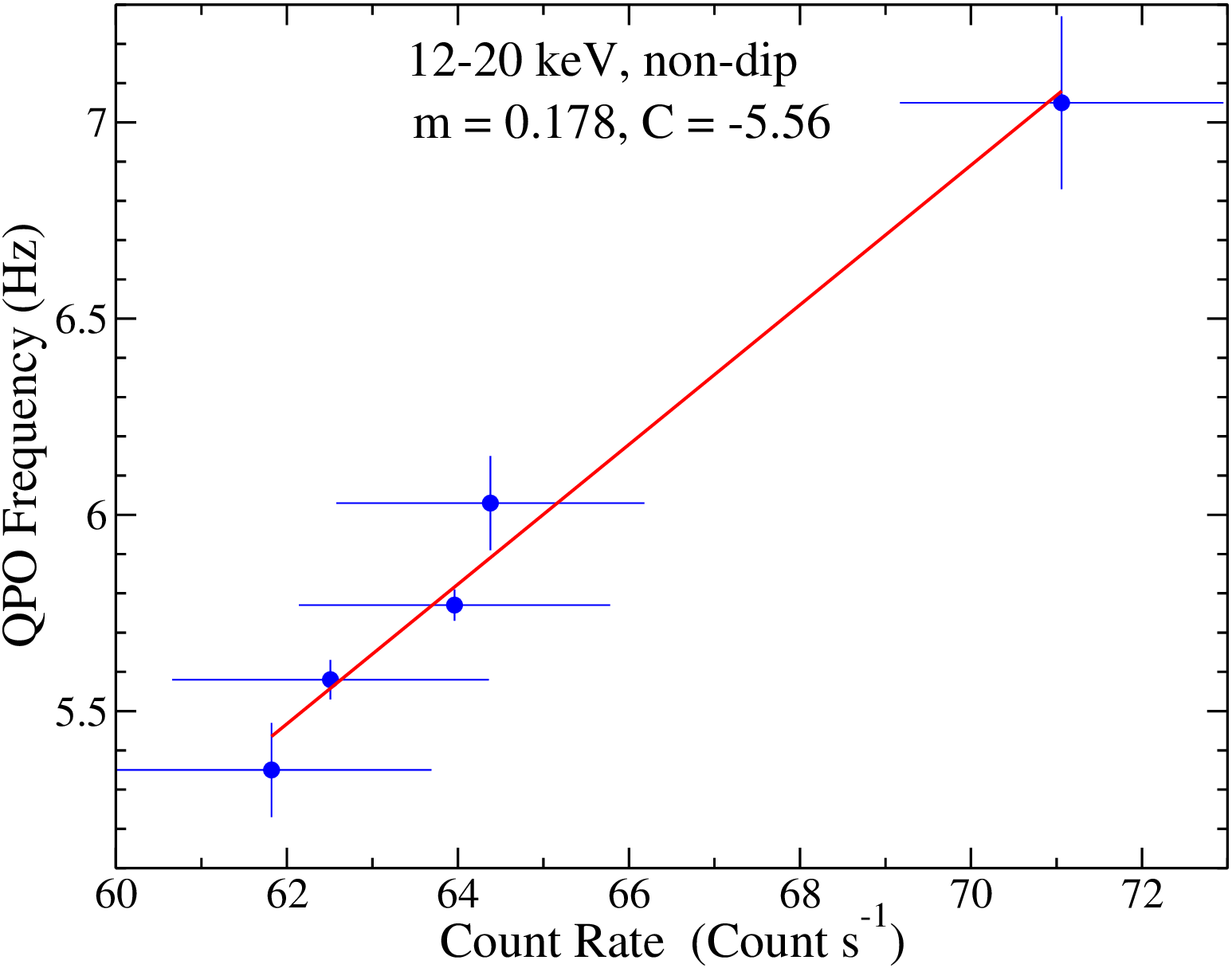}\vspace{0.2cm}
\includegraphics[width=5.5cm]{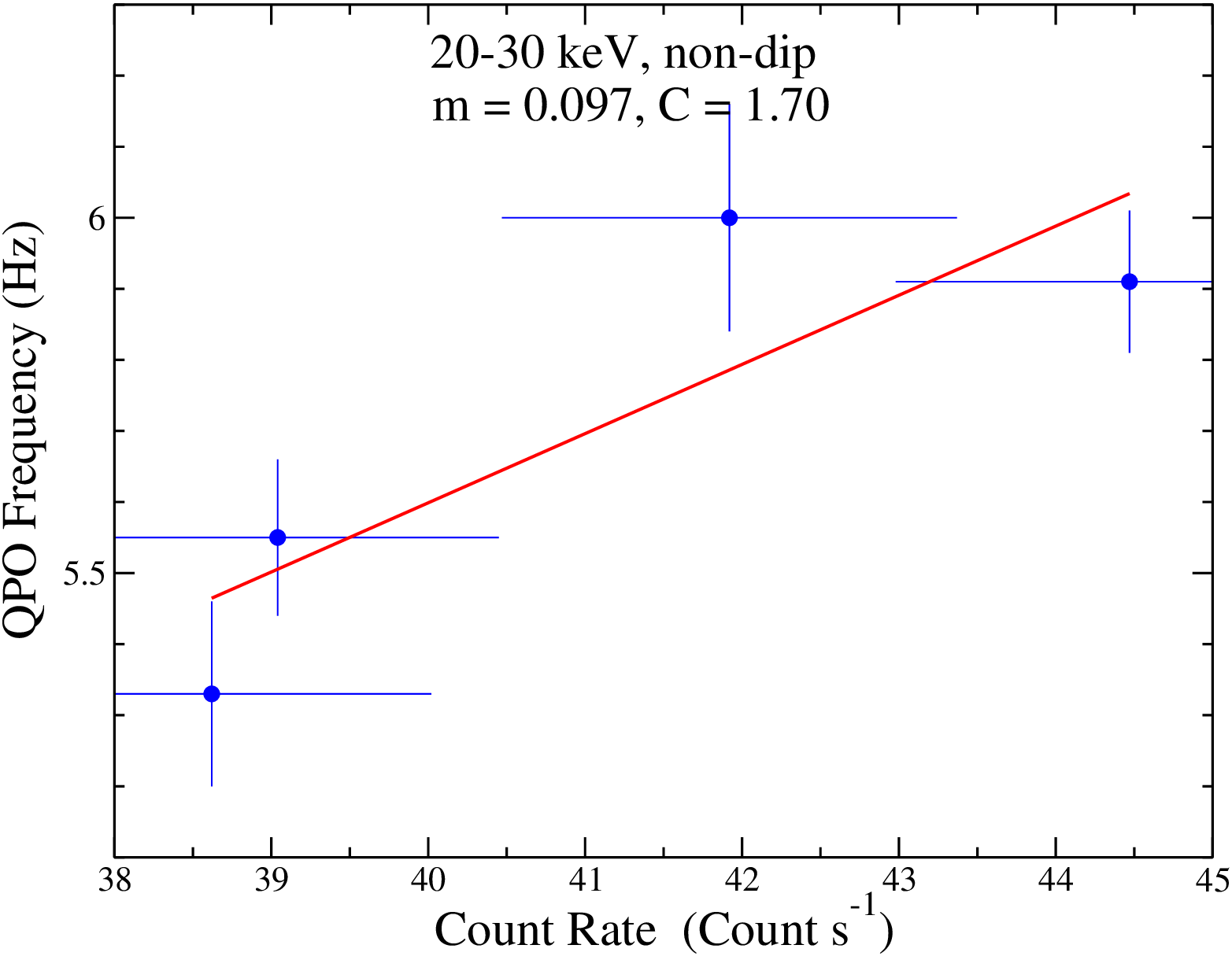}\hspace{0.2cm}
\includegraphics[width=5.5cm]{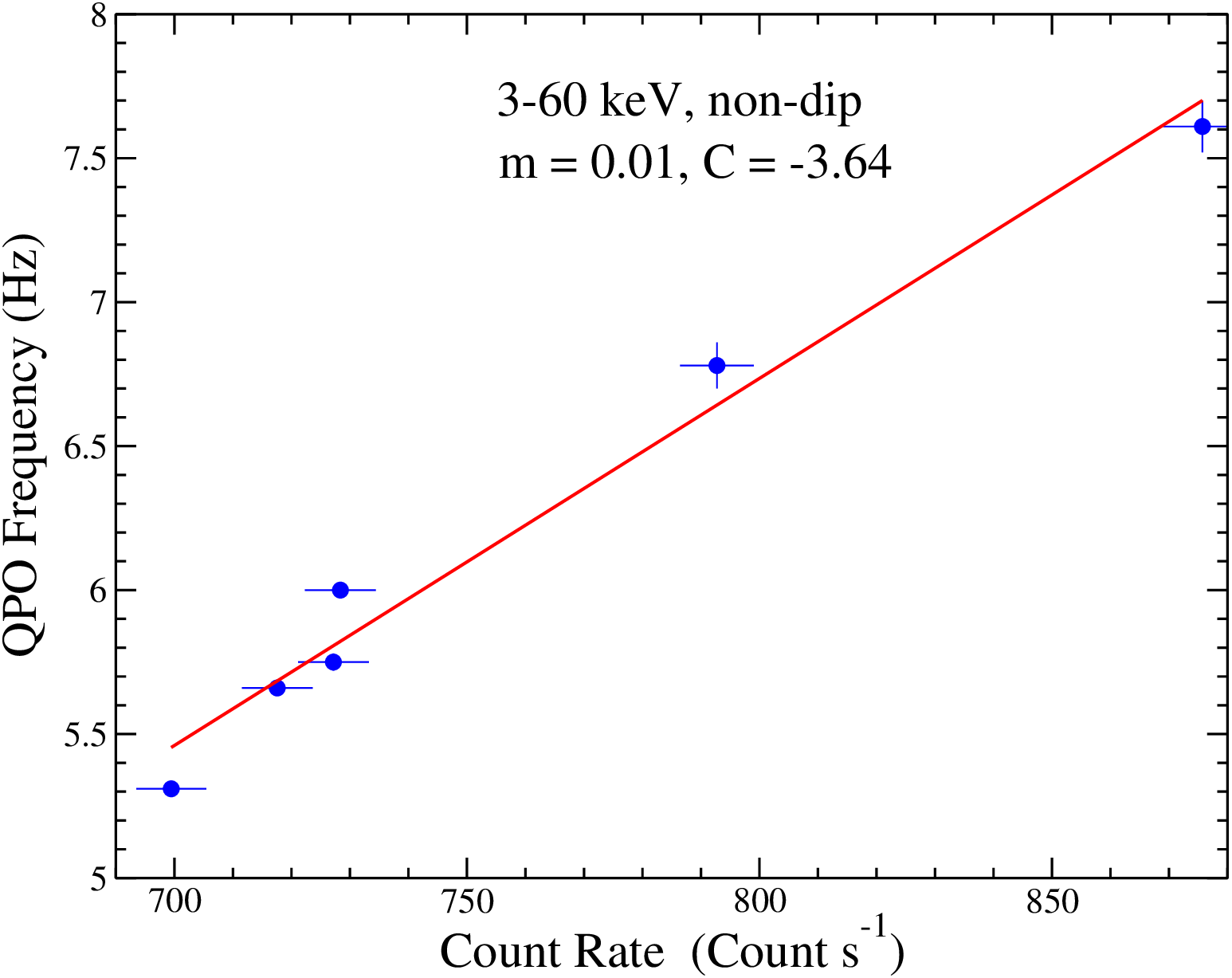}\hspace{0.2cm}
\includegraphics[width=5.5cm]{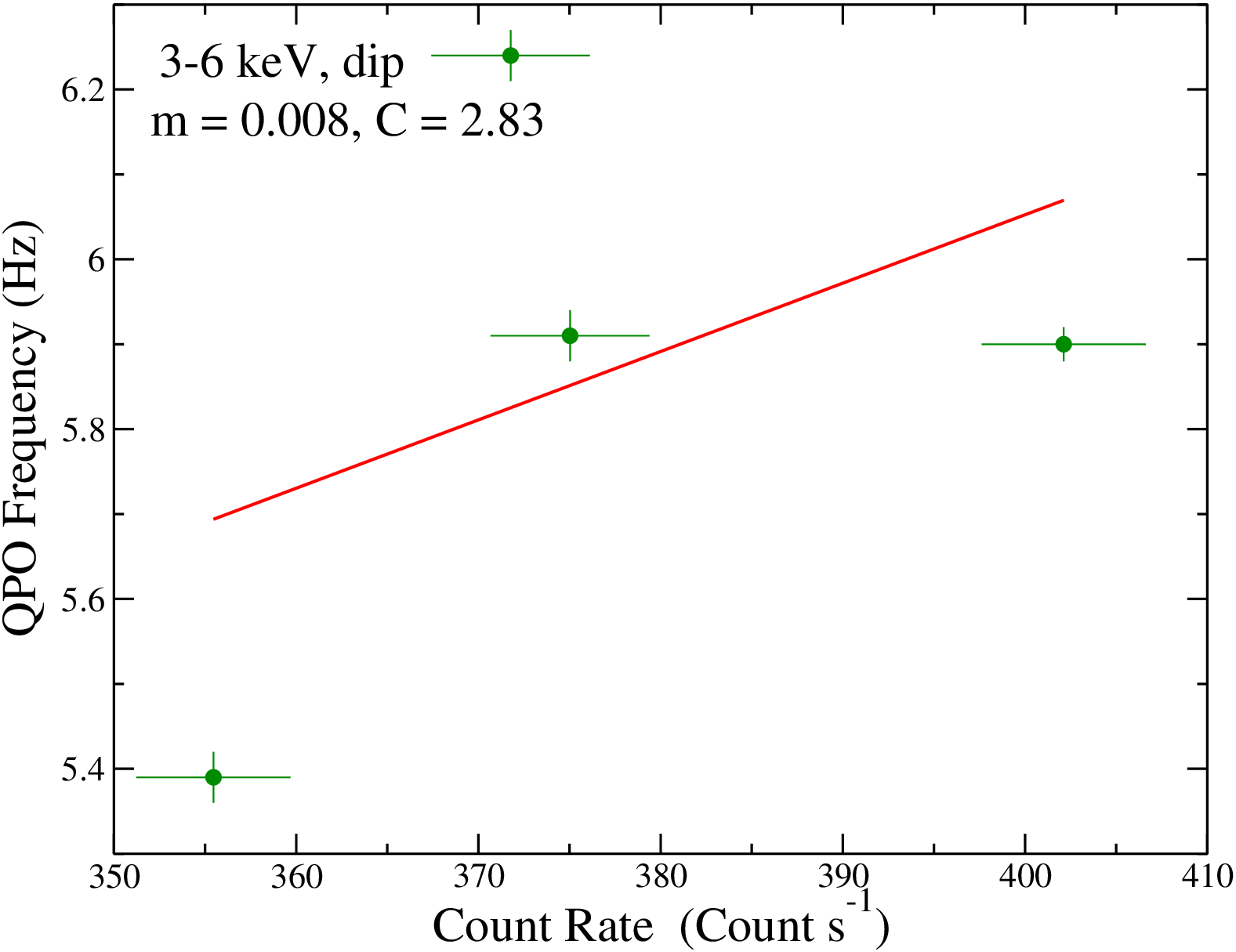}\vspace{0.2cm}
\includegraphics[width=5.5cm]{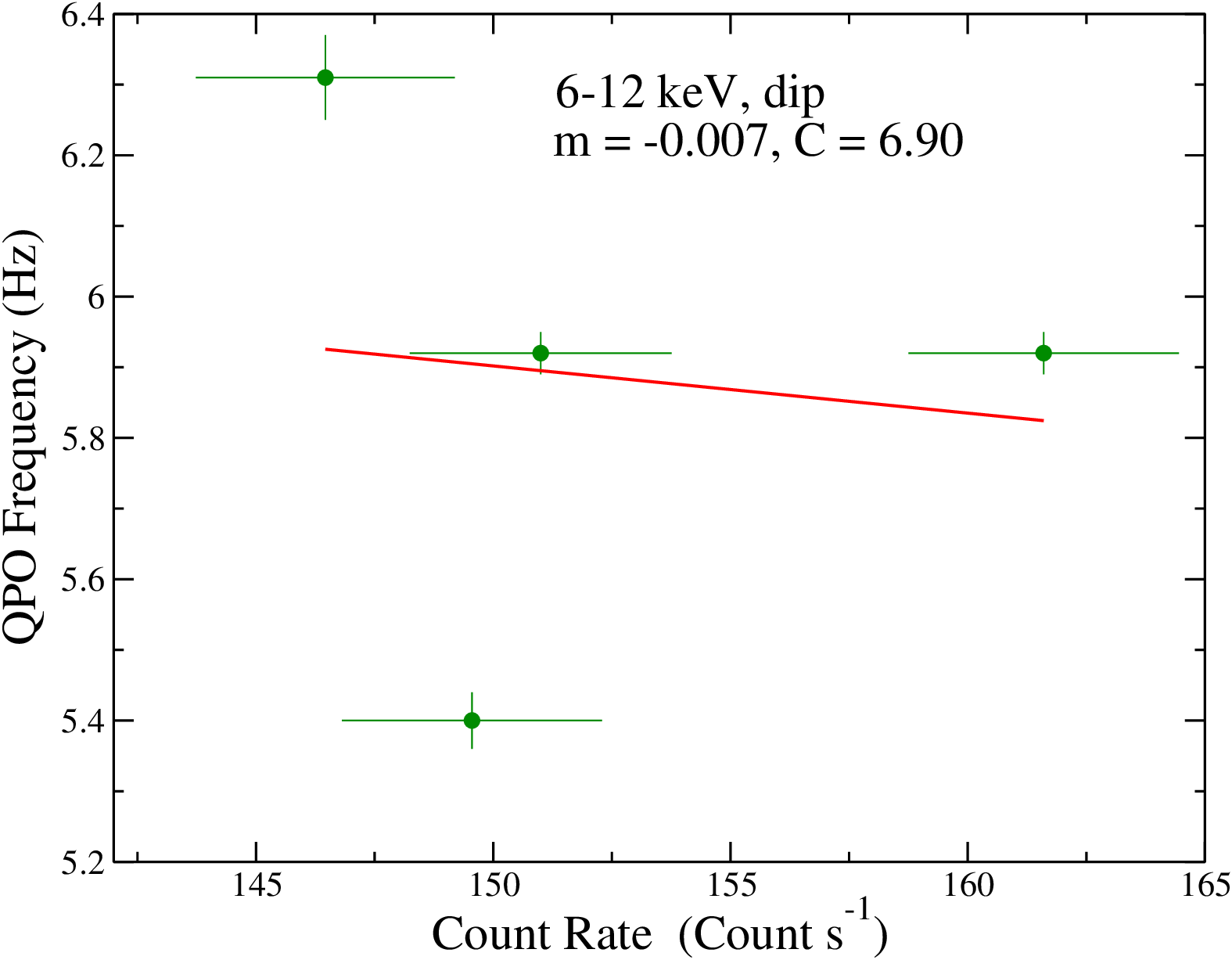}\hspace{0.2cm}
\includegraphics[width=5.5cm]{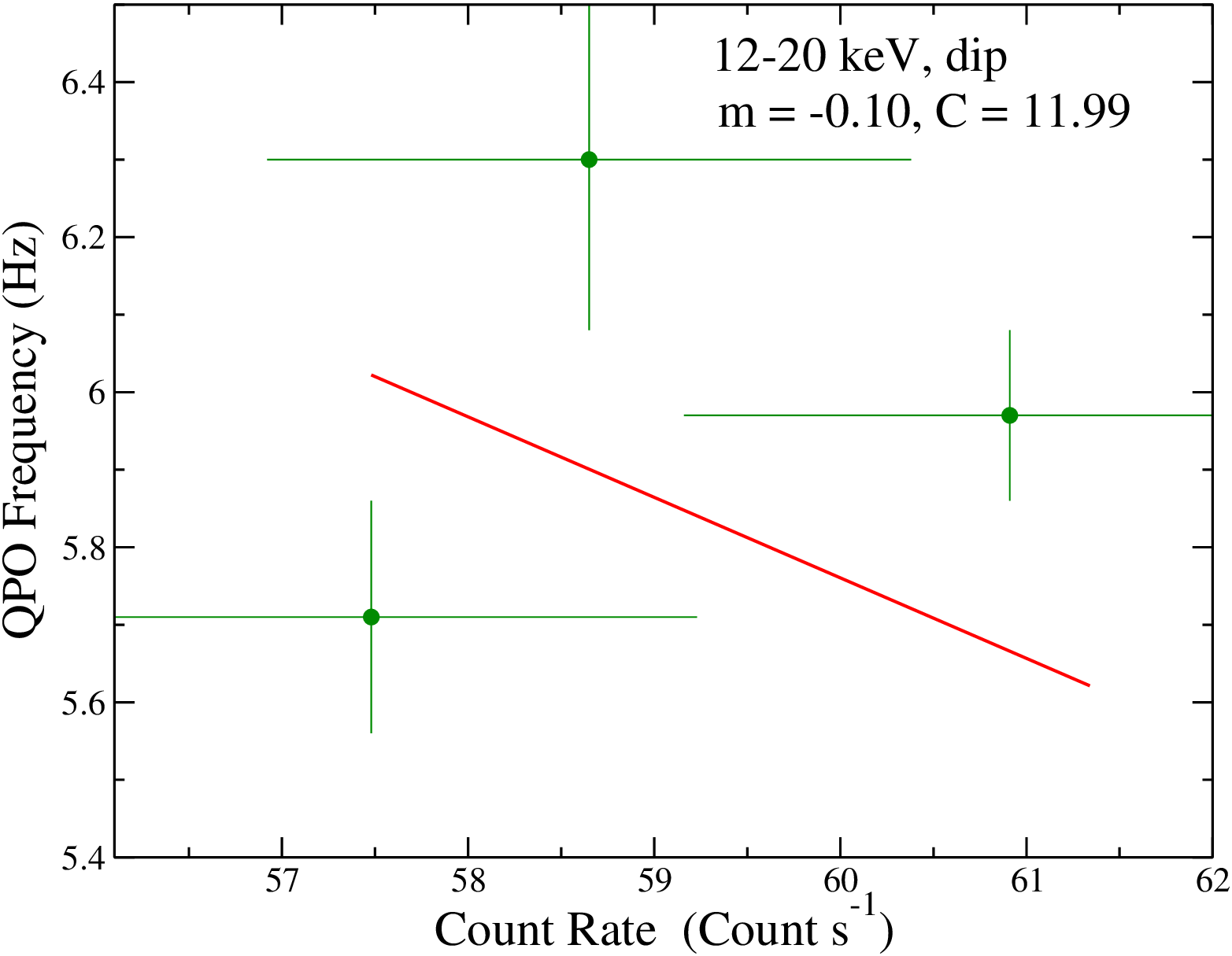}\hspace{0.2cm}
\includegraphics[width=5.5cm]{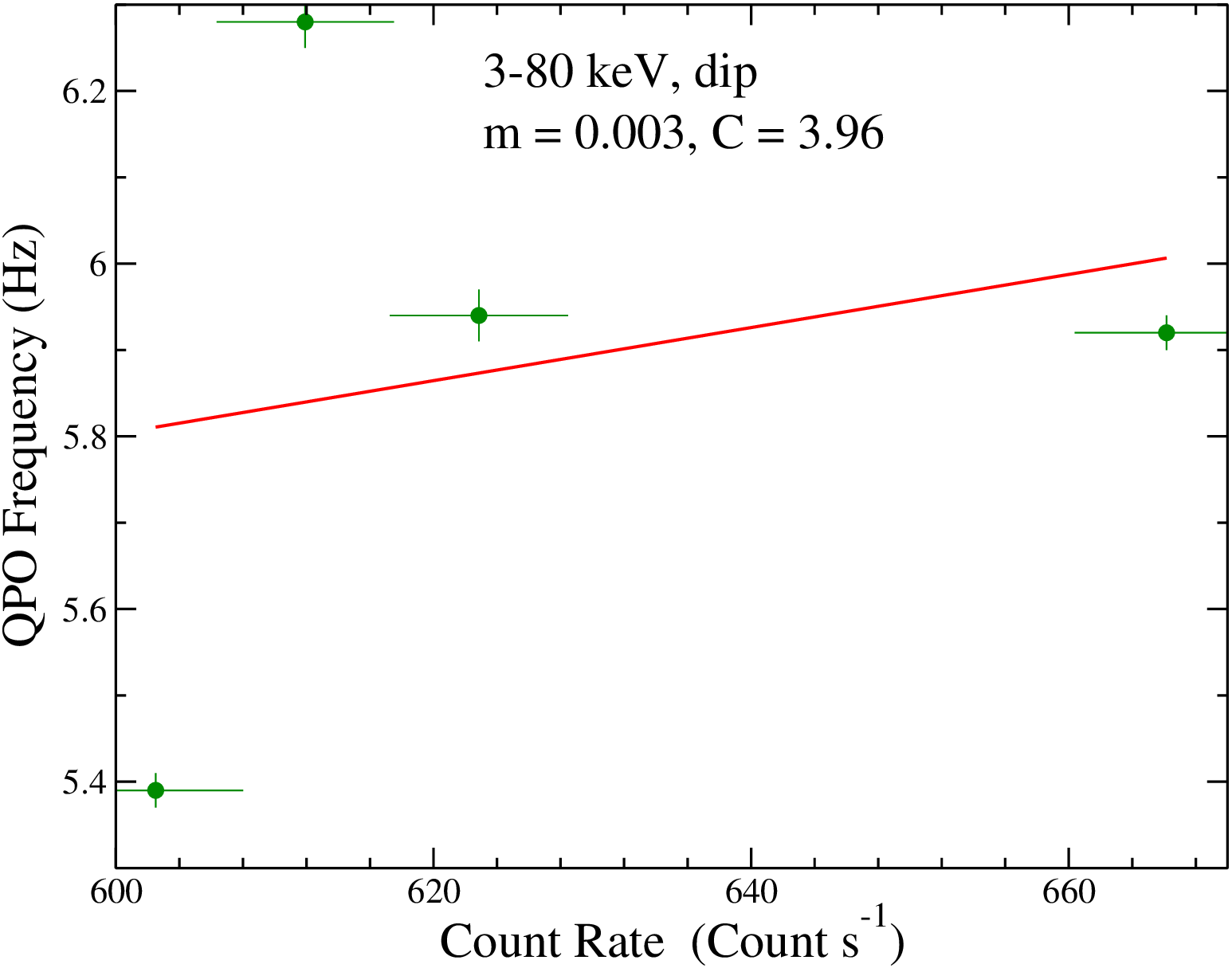}
\caption{Variation of mean count rate with QPO frequency during dip and non-dip regions are shown. The blue and green points represent non-dip and dip regions, respectively. The solid line represents the linear fit with $y=mx+c$. The value of the slope (m) and intercept (c) are mentioned in the inset of each panel.}
\label{fig:cnt-qpo}
\end{figure*}

\begin{figure}
\centering
\includegraphics[width=8.5cm]{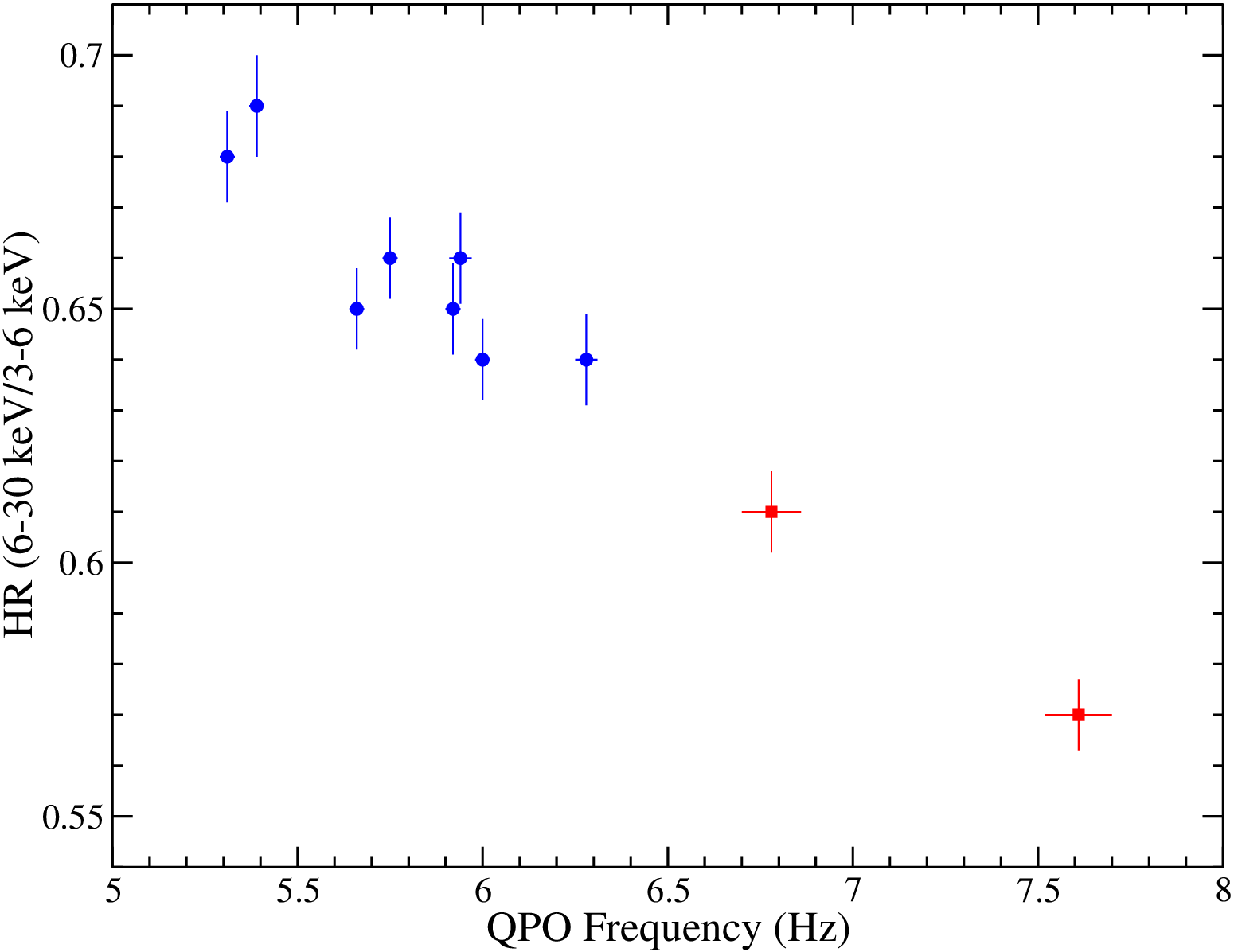}
\caption{QPO frequency is plotted as a function of hardness ratio (HR). Here, we defined HR as the ratio of mean count rate in $6-30$~keV to $3-6$~keV energy range. The blue and red points represent the data from the HIMS and SIMS, respectively.}
\label{fig:qpo-hr}
\end{figure}

Figure~\ref{fig:cnt-qpo} shows the variation of QPO frequency as a function of mean count rate for different energy bands. In this plot, we treated the dip and non-dip regions separately, as the mean count rate of the dip regions was less compared to the non-dip regions. We fitted the data points with linear regression, using $y=mx+c$. The fitted parameters (slope $m$ and intercepts $c$) are mentioned in the inset of each panel of Figure~\ref{fig:cnt-qpo}. We found that the mean count rate is strongly correlated with the QPO frequency for the non-dip regions. However, the correlation is weak for the dip regions. Figure~\ref{fig:qpo-hr} shows the QPO frequency as a function of hardness ratio. Here, we defined HR as the ratio between the mean count rates in $6-30$~keV and $3-6$~keV energy ranges. The QPO frequency is observed to be strongly anti-correlated with the HR.

\subsection{Spectral Analysis}
\label{sec:spec-analysis}
Following the procedure described in Section~\ref{sec:obs}, we extracted spectra from the SXT and LAXPC data during the dip and non-dip segments separately. The spectral analysis was carried out in $0.7-80$ keV range using combined SXT and LAXPC spectra. HEASEARC's spectral analysis software package {\tt XSPEC} v12.10 was used in spectral fitting \citep{Arnaud1996}. In the beginning, we tried to fit SXT and LAXPC spectra of each segment with absorbed power-law model. For the absorption, we used \textsc{tbabs} model with \textsc{wilm} abundance \citep{Wilms2000} and \textsc{verner} cross-section \citep{Verner1996}.

This model did not give us an acceptable fit with positive residuals below 4~keV and $6-8$~keV energy ranges. This allowed us to add a multi-colour blackbody component, \textsc{diskbb} \citep{Mitsuda1984,Makishima1986} and a \textsc{Gaussian} function at $\sim 6.4$~keV to incorporate the Fe K$\alpha$ line. We applied gain correction with fixed slope of 1 to flatten the SXT residuals at 1.8 and 2.2~keV using {\tt gain fit} command. A systematic of 3\% was added in the simultaneous fitting \citep{Antia2021}. This model in {\tt XSPEC} reads as: \textsc{tbabs*(diskbb+powerlaw+gaussian}). This combined three components model gave us an acceptable fit with $\chi^2 = 661$ for 643 degrees of freedom (dof), for the non-dip segment N1. We applied this model while fitting spectra from other non-dip segments and found that except for N6, spectra from all other regions fitted well. Though the absorption component was included in the model, additional negative residuals in soft X-rays ($<2$~keV) allowed us to add a partial absorption component \textsc{pcfabs}. However, this model did not give us a good fit as the  absorption feature was still visible in the residuals. Hence, we replaced the \textsc{pcfabs} component with the ionized absorber (\textsc{zxipcf}) component in the model. This give us an acceptable fit with $\chi^2 = 702$ for 639 dof for the segment N6.
 
We also applied the above three components model for the dips phases. However, the spectral fitting was not statistically acceptable with poor $\chi^2$. Negative residuals at soft X-rays, as in case of non-dip segment N6, allowed us to add a neutral partial absorption component \textsc{pcfabs} with above model. However, it did not give us an acceptable fit. Hence, we used ionized absorber \textsc{zxipcf} by replacing the \textsc{pcfabs} component in the model. This model fitted well the spectra of all dip phases.

\begin{figure*}
\centering
\includegraphics[width=8cm]{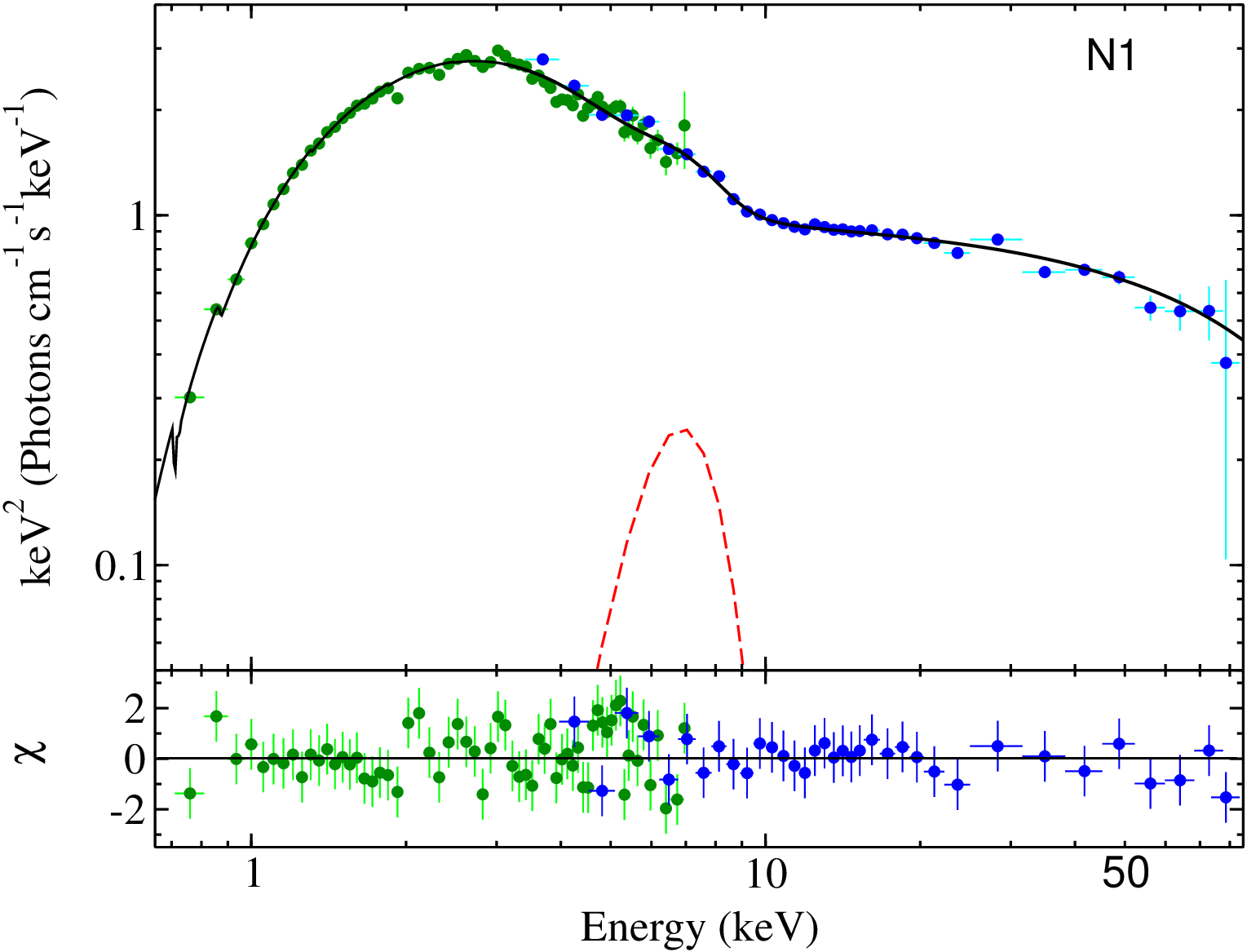}\hspace{0.5cm}
\includegraphics[width=8cm]{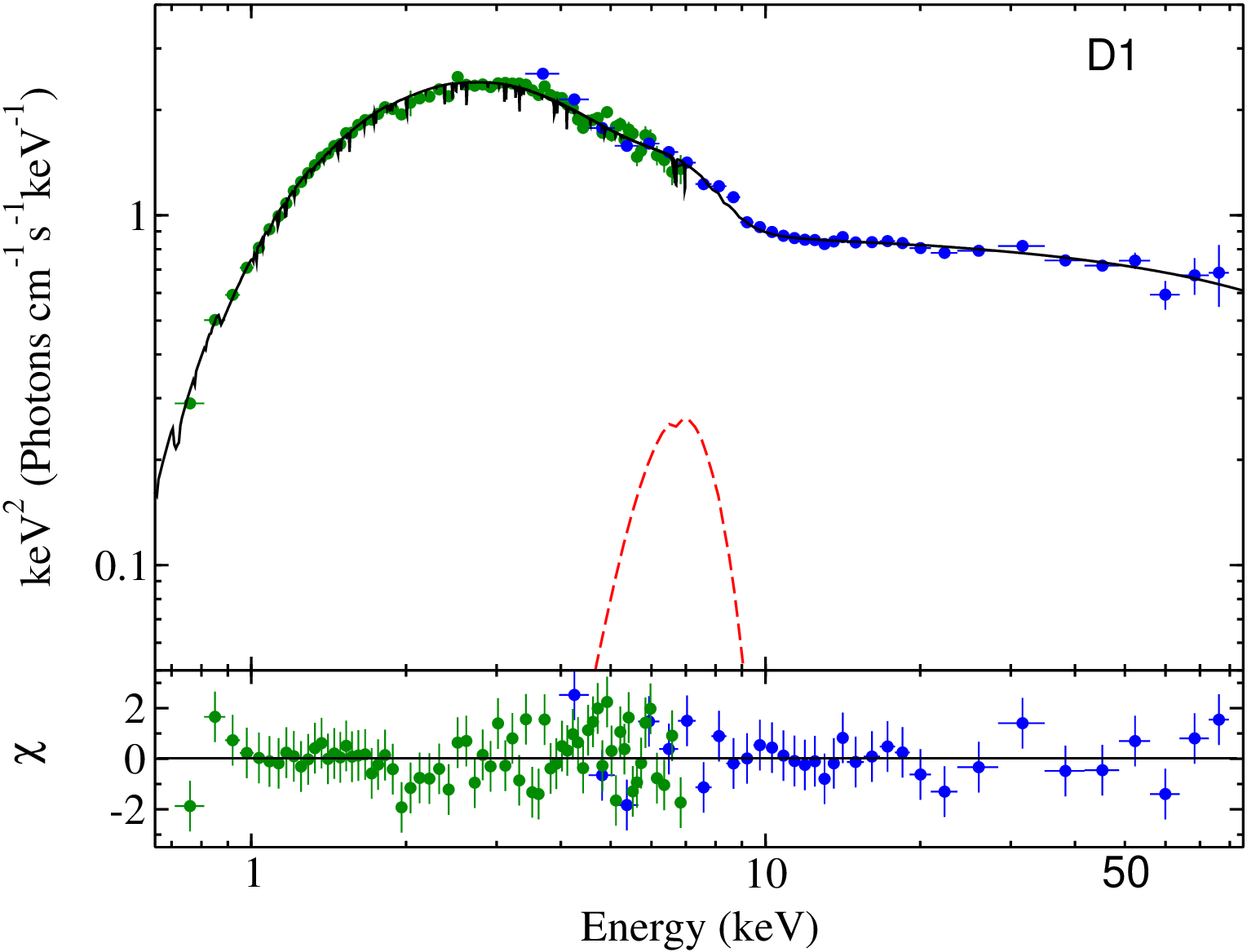}
\caption{Representative energy spectra for the non-dip (N1, left panel) and dip phases (D1, right panel) are shown along with best-fitted model components. Corresponding residuals are shown in the bottom panel. The green and blue data points indicate the SXT and LAXPC data, respectively. The black solid and red dashed lines represent the best-fit model and Fe K$\alpha$ line, respectively.}
\label{fig:spec}
\end{figure*}

\begin{figure*}
\centering
\includegraphics[angle=270,width=8.5cm]{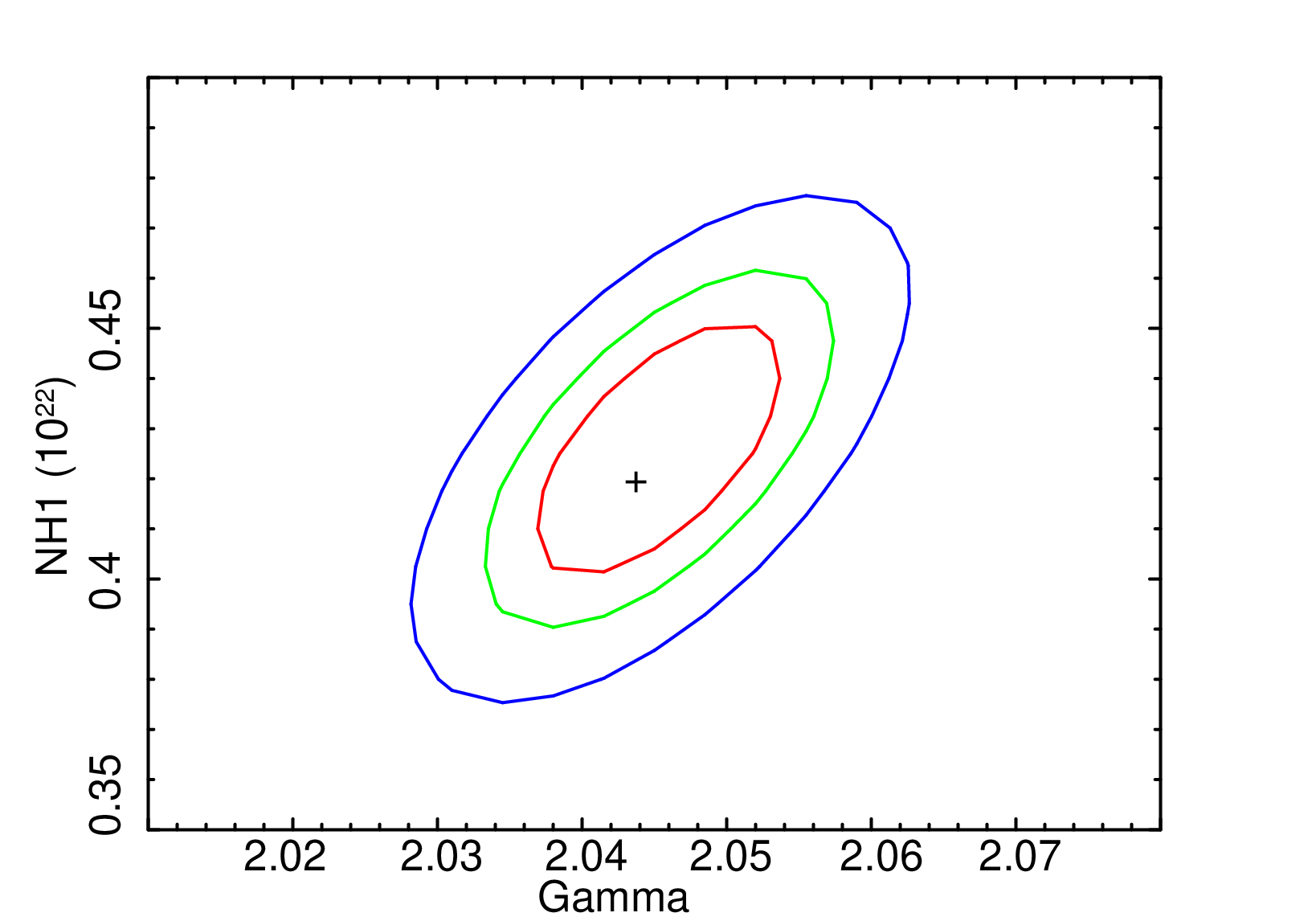}
\includegraphics[angle=270,width=8.5cm]{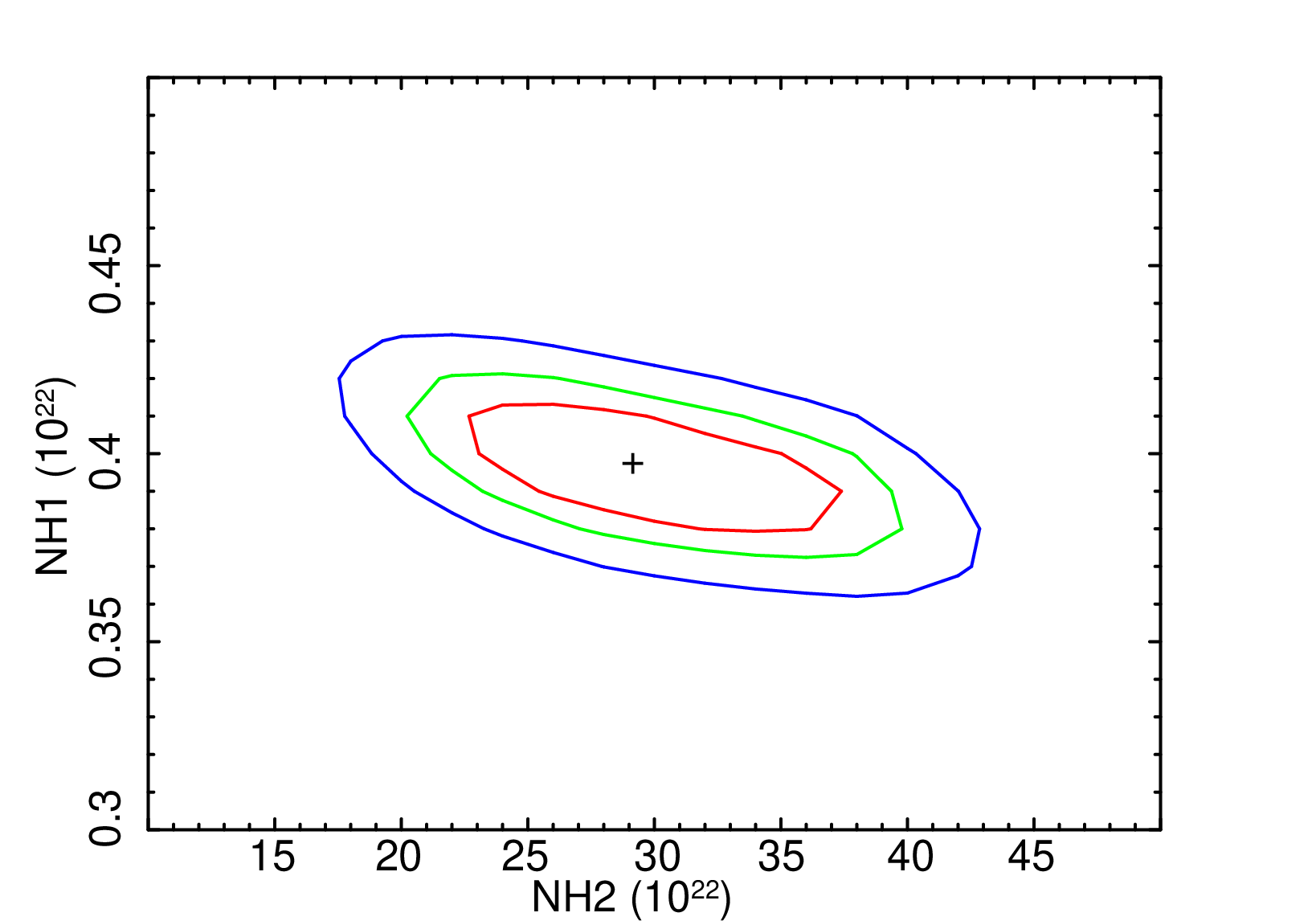}
\caption{Left panel: Confidence contours between the photon index ($\Gamma$) and neutral column density ($N_{\rm H,1}$) for the non-dip segment N1. Right: Confidence contour between the ionized column density ($N_{\rm H,2}$)  and neutral column density ($N_{\rm H,1}$) for the dip segment D1.}
\label{fig:contour}
\end{figure*}

In our analysis, we used physical Comptonization model \textsc{nthcomp} \citep{Z96,Zycki1999}, replacing phenomenological \textsc{powerlaw} model. We linked the seed photon temperature ($kT_{\rm bb}$) of the \textsc{nthcomp} model with the inner disc temperature ($kT_{\rm in}$) of the \textsc{diskbb} model. The \textsc{nthcomp} model also allowed us to estimate the hot electron plasma temperature ($kT_{\rm e}$) from the spectral fitting. Our final model reads as: \textsc{tbabs*(diskbb+nthcomp+gaussian)} for N1, N2, N3, N4 \& N5; and \textsc{tbabs*zxipcf*(diskbb+nthcomp+gaussian)} for the dip phases \& N6, respectively. Figure~\ref{fig:spec} shows the representative spectra along with the best-fitted model components from the non-dip (N1; left panel) and dip (D1; right panel) phases. Corresponding residuals are shown in the bottom panel of each spectrum. We show the confidence contour between the photon index ($\Gamma$) and neutral column density ($N_{\rm H,1}$) for N1 in the left panel of Figure~\ref{fig:contour}, whereas in the right panel, we show the confidence contour between the ionized column density ($N_{\rm H,2}$) and neutral column density ($N_{\rm H,1}$) for D1. The results obtained from the spectral analysis are tabulated in Table~\ref{tab:spec}.

\subsection{Spectral Properties}
\label{sec:spec-prop}
We obtained a good fit to the spectra from both dip and non-dip segments. The hydrogen column density of the neutral absorber was obtained to be in the range of $N_{\rm H,1} \sim 0.39-0.46 \times 10^{22}$ \pcm, which is marginally higher than the Galactic hydrogen column density, $N_{\rm H,Gal} \sim 0.25 \times 10^{22}$ \pcm~ in the direction of the source \citep{HI4PI2006}. The hydrogen column density of the ionized absorber was obtained to be $N_{\rm H,2} \sim 28 \times 10^{22}$ \pcm, during the dip phases. The partial covering factor and the ionization parameter of the ionized absorber were obtained to be $CF \sim 0.5$ and $\xi \sim 10^{3.7}$ \ecps, respectively. The ionized column density during the non-dip segment N6 was obtained to be lower than that during the dip phases, with $N_{\rm H,2} \sim 7.6 \times 10^{22}$ \pcm.

The inner disc temperature was obtained to be, $T_{\rm in} \sim 0.9$~keV during the all the segments from N1 to D4 (see Figure~\ref{fig:lxp-lc}. The $T_{\rm in}$ was observed to increase during segments N5 and reached $T_{\rm in} = 1.03 \pm 0.03$~keV during N6. We also calculated the inner disc radius ($R_{\rm in}$) from the normalization of the \textsc{diskbb} model \citep[e.g.,][]{Shimura-Takahara1995,Kubota1998,AJ2021c}, assuming the inclination angle and distance as  $i=60$\textdegree~ and $d=8$~kpc, respectively. The inner disc ($R_{\rm in}$) seemed to move towards the X-ray source during the segments N5 \& N6 with $R_{\rm in}$ decreasing from $\sim 54-59$~km during N1 -- D4 to $\sim$ 46~km during N6. The photon index was also observed to be $\Gamma \sim 2$ during N1 -- D4 which  increased to $\Gamma = 2.12 \pm 0.03$ during N6. The Compton cloud temperature ($kT_{\rm e}$) did not show any significant changes during all the segments and was found to be variable in $30-35$~keV range. We also calculated the optical depth ($\tau$) of the Compton corona using the following equation \citep{T94,Z96},

\begin{equation}
\tau \approx \sqrt{\frac{9}{4}+\frac{m_{\rm e}c^2}{kT_{\rm e}}\frac{3}{(\Gamma -1)(\Gamma + 2)}} - \frac{3}{2}.
\label{eqn:tau}
\end{equation}

\begin{figure}
\centering
\includegraphics[width=8.5cm]{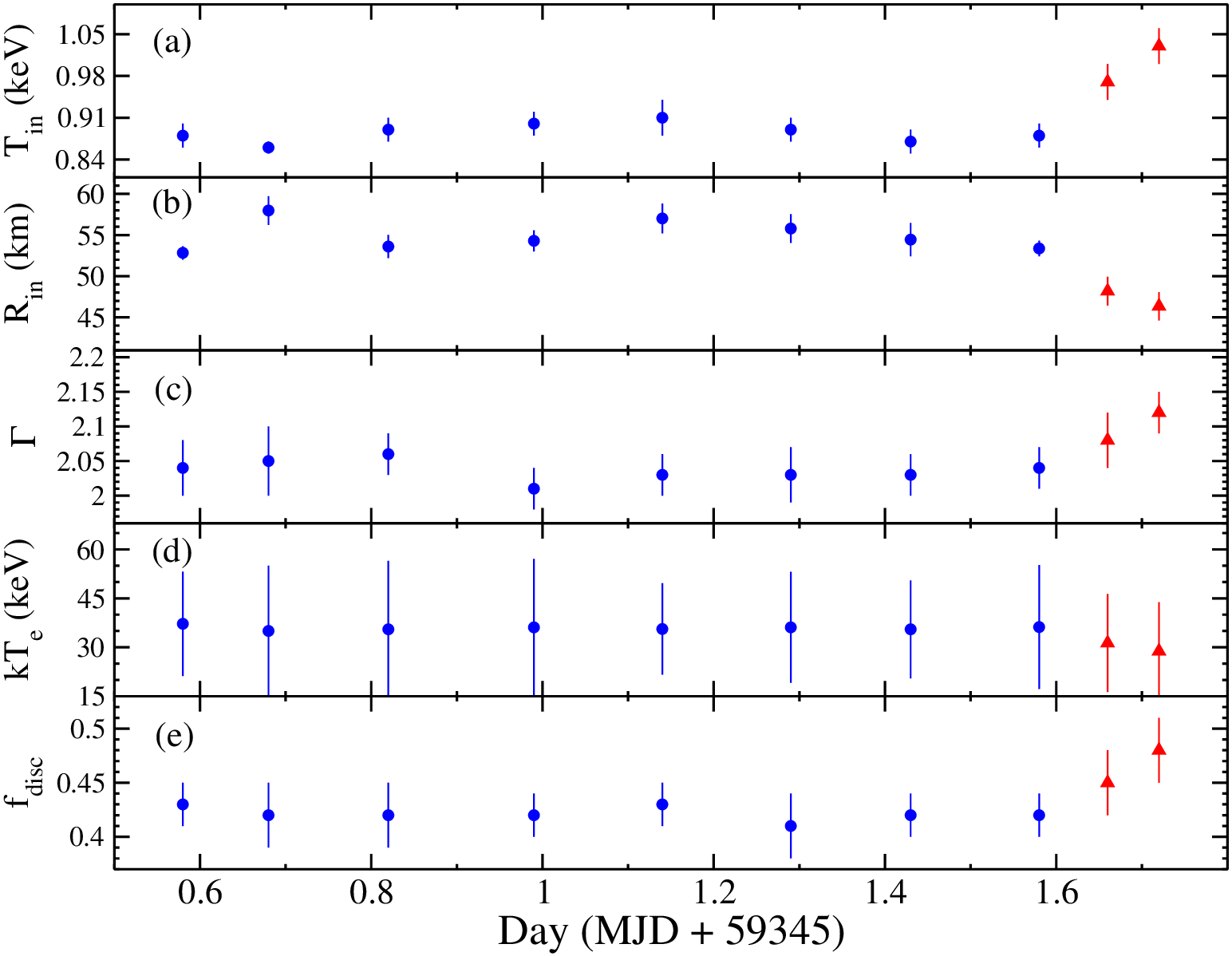}
\caption{Variation of the spectral parameters: (a) inner disc temperature ($T_{\rm in}$) in keV, (b) inner disc radius ($R_{\rm in}$) in km, (iii) photon index ($\Gamma$), (iv) Compton cloud temperature ($kT_{\rm e}$) in keV, and (v) thermal emission fraction ($f_{\rm disc}$) are shown from top to bottom panels, respectively. The blue circles and red triangles represent the data from the HIMS and SIMS, respectively. $R_{\rm in}$ is calculated assuming the distance $d=8$~kpc and inclination angle $i=60$\textdegree.}
\label{fig:evo-spec}
\end{figure}

The optical depth was estimated to be, $\tau \sim 2$ during segments from N1 to D4, which increased to $\sim 2.2$ during the segment N6. Although, within the uncertainties, the optical depth was estimated to be constant. During our analysis, we obtained the width ($\sigma$) of the Fe K$\alpha$ line to be $\sigma \sim 1-1.3$~keV with the equivalent width of EW$\sim 0.5-0.7$~keV. We also calculated the thermal emission fraction as, $f_{\rm disc} = F_{\rm disc}/F_{\rm tot}$, where $F_{\rm disc}$ and $F_{\rm tot}$ are the unabsorbed thermal flux and unabsorbed total flux in $0.1-100$~keV range. We found that the $f_{\rm disc} \sim 0.4$ during segments N1 to D4 which then increased to $\sim 0.48$ during N6. The evolution of the spectral parameters are shown in Figure~\ref{fig:evo-spec}. The variation of (a) inner disc temperature ($T_{\rm in}$) in keV, (b) inner disc radius ($R_{\rm in}$) in km, (iii) photon index ($\Gamma$), (iv) Compton cloud temperature ($kT_{\rm e}$) in keV, and (v) thermal emission fraction ($f_{\rm disc}$) are shown in top to bottom panels of Fig~\ref{fig:evo-spec}.

\section{Discussion}
\label{sec:dis}
We studied the timing and spectral properties of MAXI~J1803--298 in $0.7-80$~keV  energy range, using the SXT and LAXPC data from the \astrosat~ observations during the recent X-ray outburst. For the timing studies, we used the LAXPC data in different energy bands. For the spectral studies, we used the combined data in $0.7-7$~keV and $3-80$~keV ranges from the SXT and LAXPC instruments.

\subsection{Absorption Dips}
We found four periodic absorption dips with a periodicity $P = 7.02 \pm 0.18$~hr, in the $3-80$~keV LAXPC light curve. Although, different energy bands yield different periods (see Table~\ref{tab:LS}), they are consistent within uncertainties. \nustar~ observation show the periodicity of $7.18\pm0.92$~hr in the $3-78$~keV energy range, which is consistent with the findings from the $3-80$~keV LAXPC light curve. As the average source flux during the dip was $\sim 7.9 \times 10^{-9}$ \eps~ (in $3-80$~keV range), about $\sim 10$ \% less than the flux during the non-dip phases, these dips in the light curve can not be due to the eclipse of the X-ray source by the binary companion. Usually, the absorption dips are observed at the orbital period of the system and understood to be caused due to the obscuration by the material in the buldge (thickened material) of the outer accretion disc \citep{Frank1987}. The buldge is known to be formed due to the interaction of the outer accretion disc with the in-flowing stream of gas from the companion star \citep[e.g.,][]{Walter1982,White1982a}. The period of the absorption dips are identified as the orbital period of the system, following other X-ray dippers \citep{Zurita2008,Kuulkers2013}.

In our analysis, we found that the hard X-ray photons ($>6$~keV) were less affected due to absorption during the dip phases (see Figure~\ref{fig:lxp-lc}). The spectral analysis showed that a neutral absorber with column density, $N_{\rm H,1} \sim 0.4 \times 10^{22}$ \pcm~ was present during dip and non-dip phases. An additional ionized absorber was detected during the dip phases, indicating that the dips are associated with the obscuration due to buldge at the outer accretion disc. The ionized absorber had about $\sim 70$ times higher column density ($N_{\rm H,2} \sim 28 \times 10^{22}$ \pcm) compared to the neutral absorber.

The ionization parameter ($\xi$) and column density ($N_{\rm H,2}$) of the ionized absorber are approximately constant during our observation period. The ionization is observed to be $\xi \sim 10^{3.7}$ \ecps. This value of ionization is comparable to that reported in other BHXRBs, e.g., GRO~J1655--40 \citep{Diaztigo2007}, GRS~1915+105 \citep{Kotani2000}. As we obtained $\xi$ and $N_{\rm H,2}$ from the spectral analysis, we estimated the incident luminosity on the ionized absorber. By definition, the ionization parameter is given by, 

\begin{equation}
\xi = \frac{L_{\rm X}}{n_{\rm H}R^2}= \frac{L_{\rm X}}{N_{\rm H}R}\frac{\Delta R}{R}.
\end{equation}

Here, $n_{\rm H}$, $\Delta R$ and $L_{\rm X}$ represent the hydrogen number density, length of the absorber and incident X-ray luminosity, respectively. In BHXRBs, a typical ionized absorber is found to be located at $R \sim 10^{4-6}$ km, when $\Delta R/R$ is assumed to be 1 \citep[e.g.;][]{Shidatsu2014}. Assuming the absorber to be located at the outer disc in \source, the location of the absorber is $R \sim 4.2-4.8 \times 10^5$ km (see Section~\ref{sec:bin-par}). We can estimate the X-ray luminosity ($L_{\rm X}$) during the dip phase by assuming $\Delta R/R$=1, and the ionized absorber is located at the outer disc. The luminosity is estimated to be $L_{\rm X} \sim 6-8 \times 10^{37}$ \eps. The observed value is consistent with the luminosity of typical BHRXB during the outburst phase. From this, the source is found to be accreting with $6 - 10$\% of Eddington luminosity.

\subsection{Estimation of the Binary Parameters}
\label{sec:bin-par}
The presence of periodic absorption dips allows us to derive the orbital period of the system. The presence of dips also suggests the binary to be a high inclination system, i.e. $i>60$\textdegree. The dips in the light curve which are not due to eclipse of the X-ray source, allow us to constrain the upper limit on the inclination i.e., $i<75$\textdegree~ \citep{Frank1987}.

The mass function of the binary system is given by, 
\begin{equation}
f(M) = \frac{(M_{\rm BH} \sin~i)^3}{(M_{\rm BH} + M_2)} = \frac{(M_{\rm BH} \sin^3~i)}{(1+q)^2} = \frac{P~K_2^3 }{\pi G},
\label{eqn:fm}
\end{equation}

where $M_{\rm BH}$, $M_2$, $P$, $i$, $K_2$ and q are the BH mass, companion mass, orbital period, inclination of the binary system, velocity of the companion and the mass ratio $q = M_2 / M_{\rm BH}$, respectively.

In LMXBs, most of the optical emission comes from the outer region of the accretion disc, mainly due to X-ray irradiation \citep{vanparadijs1994,vanparadijs1996,Shahbaz1996,Tetarenko2021}. \source~ was observed in the optical wavebands with the Robert Stobie Spectrograph on the Southern African Large Telescope (SALT), Gran Telescopio Canarias (GTC) and Very Large Telescope (VLT). The optical spectra from these observations are found to be consistent with the spectra of LMXBs \citep{Buckley2021,Mata2022}. During the outburst, a double peaked H$\alpha$ line was observed. Between May 5, 2021 and July 2, 2021, the peak-to-peak separation of H$\alpha$ line is observed to be $\sim 500-700$ \kms \citep{Mata2022}. Assuming that the H$\alpha$ line is arising from the outer accretion disc, we have a conservative estimation of rotational velocity of the outer disk as, $v_{\rm d} \sin i \sim 500-700$ \kms.

We now estimate the radial velocity of the donor star ($K_2$) based on the relation, $v_{\rm d}/K_2 \approx 1.1-1.25$ \citep{Orosz1994,Orosz1995,Shaw2016}. From this, the radial velocity is estimated to be, $K_2 \approx 430-730$ \kms, assuming $i = 60$\textdegree -- 75\textdegree. We further estimated the mass function of the system as $f(M) = 2.4-11.7~M_{\odot}$ using Equation~\ref{eqn:fm}. From the FWHM of H$\alpha$ line, \citet{Mata2022} estimated the radial velocity as, $K_2 = 410-620$ \kms~ at 95\% confidence. Using this, the mass function is estimated to be, $f(M) = 2.1-7.2~M_{\odot}$.

In order to calculate the BH mass, we need the information of the mass of the donor star ($M_{\rm 2}$) or the mass ratio ($q = M_{\rm 2}/M_{\rm BH}$). The mass ($M_{\rm 2}$) and radius ($R_{\rm 2}$) of the donor star can be calculated by assuming the accretion via Roche-lobe overflow. Considering, the the secondary star filled its Roche-lobe and assuming the mass ratio $q < 0.8$, the mean density of the lobe-filling star can be estimated from the orbital period solely \citep{Eggleton1983}. The mean-density of the lobe filling star is \citep{Paczynski1971,Eggleton1983,Frank2002}, 

\begin{equation}
\bar{\rho} = \frac{3M_{\rm 2}}{4 \pi R_{\rm 2}^3} \approx \frac{3^5 \pi}{8G P^2} \approx 110~P_{\rm hr}^{-2}~{\rm g}~{\rm cm}^{-3}.   
\label{eqn:density}
\end{equation}

The mean density is obtained from the above equation is $\bar{\rho} \sim 2.2$ \gpc. Re-writing Equation~\ref{eqn:density}, one can obtained the following relation to calculate the mass and the radius of the donor star \citep{Frank2002}, 

\begin{equation}
M_{\rm 2} \approx 0.11~P_{\rm hr}~{\rm M_{\odot}},    
\label{eqn:m2}
\end{equation}
and,
\begin{equation}
R_{\rm 2} \approx 0.11~P_{\rm hr}~{\rm R_{\odot}}.    
\label{eqn:r2}
\end{equation}

Using Equation~\ref{eqn:m2} and Eqn~\ref{eqn:r2}, the mass and the radius of the donor star are estimated to be, $M_{\rm 2} = 0.77 \pm 0.02~M_{\odot}$ and $R_{\rm 2} \sim 0.77 \pm 0.02~M_{\odot}.$

We also tried to calculate the mass and radius of the donor star from the empirical mass-radius relationships for the donor stars in the Cataclysmic Variable (CV) binaries, using the following relations \citep{Smith1998}, 

\begin{equation}
M_{\rm 2} = (0.126\pm 0.0011) P_{\rm hr} - (0.11 \pm 0.04).
\label{eqn:m2cv}
\end{equation}
and 
\begin{equation}
R_{\rm 2} = (0.117\pm 0.004) P_{\rm hr} - (0.041 \pm 0.018).
\label{eqn:r2cv}
\end{equation}

We estimated the mass and radius of the donor star as $M_{\rm 2} = 0.77 \pm 0.05~M_{\odot}$ and $R_{\rm 2} = 0.78 \pm 0.04~R_{\odot}$, respectively which are consistent with the above findings. Using the same relation, \citet{Kuulkers2013} estimated the mass and radius of the companion in a black hole X-ray binary MAXI~J1659--152. \citet{Warner1995a} derived the mass-period relation for the CV as $M_{\rm 2} = 0.065~P^{5/4}$. From this, the mass of the companion star would be $M_{\rm 2} \sim 0.74 \pm 0.03~M_{\odot}$, which is consistent with the above findings. From the above estimation of the mass, radius and density, the companion star could be a K-type star if it is a main-sequence star \citep{Keenan1989,Pecaut2013}. This means that \source~ is an LMXB as suggested from the optical observations \citep{Hosokawa2021}.

We tried to estimate the mass of the BH from the estimated mass function and mass of the companion star. {The mass function is estimated to be $f(M) = 2.1-7.2~M_{\odot}$}. From this, the mass of the BH is estimated to be, $M_{\rm BH} \sim 3.5-12.5~M_{\odot}$, for the inclination angle $i = 60$\textdegree -- 75\textdegree~ and mass of the donor star as $M_{\rm 2} = 0.77 \pm 0.05~M_{\odot}$. \citet{Mata2022} also estimated the mass of BH as $M_{\rm BH} \sim 3-10~M_{\odot}$, assuming $i>65\textdegree$ and $0.01<q<0.2$. This is consistent with our estimation.

We further estimated the size of the binary system and the outer disc radius. The binary separation or size is given by \citep{Eggleton1983},
\begin{equation}
a=3.5\times10^{10} (M_{\rm BH})^{1/3} (1+q)^{1/3} P_{\rm hr}^{1/3}~{\rm cm}.
\label{eqn:bin-size}
\end{equation}

For the estimated BH mass range ($M_{\rm BH} \sim 4.8-9$ $M_{\odot}$), the binary separation is obtained to be $a \sim (1.2-1.4) \times 10^6$ km. The binary size is clearly on the lower side comparing other BHXRBs. The Roche lobe for the secondary is given by \citep{Eggleton1983},

\begin{equation}
R_{\rm 2,L} = \frac{0.49q^{2/3}a}{0.6q^{2/3}+ln(1+q^{1/3})}.
\label{eqn:rl}
\end{equation}

One can estimate the Roche lobe for the primary by replacing `q' with `$q^{-1}$'. The Roche-lobe for the primary is estimated to be $R_{\rm 1,L} \sim (6.3-7.2) \times 10^5$~km. Now, assuming, the outer radius of the accretion disc ($R_{\rm out}$) as 2/3 of $R_{L}$, the outer radius of the accretion disc is estimated to be $R_{\rm out} \sim 4.2-4.8 \times 10^5$~km.

\subsection{Spectral State Transition}
\label{sec:state}
An outburst is believed to be triggered by the sudden enhancement of the viscosity at the outer edge of the disc \citep{Ebisawa1996,Bhowmick2021} or disc instability \citep[e.g.,][]{Lasota2001}. At the beginning of the outburst, the source is believed to be in the LHS when the disc is truncated at a large distance with a dominance of Comptonized emission \citep{RM06}. As the outburst progresses, the disc moves towards the BH, causing the disc emission to dominate over the Comptonized emission and the source moves through the HIMS, SIMS, and HSS. In the declining phase, the source evolves through the SIMS, HIMS and LHS and the disc moves outwards.

During the present outburst of \source, the source was in the LHS when the outburst started. Then, it moved through the IMS and HSS before entering to the declining phase. Throughout the outburst, the HR was found to be constant. This did not allow us to infer the spectral state classification explicitly. A detail study of the spectral and timing properties in a daily basis is required which is out of scope of the present work.

During the first \astrosat~ observation, the evolution of the timing and spectral parameters (presented in Section~\ref{sec:res}) indicates the changes in the accretion geometry during the observation period. From the spectral and timing properties, it is clear that \astrosat~ observed the source during the state transition from HIMS to SIMS. During the first eight segments (N1 to D4 in Figure~\ref{fig:lxp-lc}), the source was in the HIMS with nearly constant flux. An increasing flux (count rate) was observed during the segment N5 during which the source entered to the SIMS and remained there till the end of our observation. Additionally, the hardness ratio (HR1) decreased, indicating the dominance of the soft photons in the SIMS. 

Evolving type-C QPOs are known to be detected in the HIMS and LHS, whereas the sporadic type-A or type-B QPOs may be observed in the SIMS \citep{Nandi2012}. However, no QPOs are observed in the HSS \citep{Belloni2005,RM06}. The oscillation of the Compton cloud is believed to be responsible for the QPOs \citep{Molteni1996,Titarchuk1998,Cabanac2010}. It is well established that the Comptonized photons are responsible for the variabilities observed in the PDS and the light curves \citep{CM2000,vanderklis2004}. In general, strong variabilities are observed in the LHS with fractional rms amplitude of $\sim 20 - 40$\%. In the HIMS and SIMS, the fractional rms amplitudes of $\sim 10 - 20$\% and $\sim 5 - 10$ \% are seen, respectively. In the HSS, however, weak variabilities with fractional rms amplitude of $<5$\% are observed \citep{vanderklis1994,vanderklis1997,Belloni2005}. During our observation, we detected evolving type-C QPOs with strong variabilities with fractional rms amplitude of $\sim 15-20\%$. This indicated that the observation was made during the intermediate state, i.e. HIMS \& SIMS.

We also calculated the rms in different energy bands and constructed the rms spectra. The rms was observed to be maximum in $6-12$~keV energy band with the rms $\sim 15-20 \%$ and decreased in the higher energy bands. In general, flat and inverted rms spectra are observed in the LHS and HSS, respectively \citep{Gierlinski2005}. In the intermediate states (HIMS and SIMS), the rms spectra are generally characterised with two slopes along with the peak appearing at certain energy band \citep{Gierlinski2005}. The observed rms spectra (in present work) are similar to that generally observed in the intermediate state \citep{Gierlinski2005,Shaposhnikov2010,Rout2021}. 

The QPO frequency is found to vary in the range of $5.31\pm 0.02 - 6.28 \pm 0.03$~Hz during the segments from N1 to D4, which then increased to $\sim 7.61 \pm 0.09$~Hz during the segment N6. The Q-factor and rms (QPO rms and broadband rms) also decreased during N5 and N6, indicating a different spectral state. The observed spectral properties also supported the state transition. The inner disc temperature was observed to be $T_{\rm in} \sim 0.9$~keV during the HIMS, while it increased to $\sim 1$~keV during the SIMS. The photon index was also observed to become soft during the SIMS. The thermal emission fraction ($f_{\rm disc}$) was observed to increase in the SIMS from 0.42 to 0.48.

\subsection{Accretion Geometry}
\label{sec:geom}
It is suggested that the QPO frequency ($\nu_{\rm QPO}$) would vary with the Compton cloud boundary ($X$), as $\nu_{\rm QPO} \sim 1/X^{3/2}$ \citep{CM2000,SKC2008}. Thus the observed variation in the QPO frequency indicated an evolving Compton cloud. In the HIMS (segments N1 to D4), the QPO frequency varied in the range of $5.31\pm 0.02 - 6.28 \pm 0.03$~Hz. The QPO frequency increased in the SIMS (N5 \& N6) to $\nu_{\rm QPO} \sim 7.61 \pm 0.09$~Hz, which indicated a contracting Compton corona. The evolution of the Compton cloud is also evident from the observed changes in the optical depth. The corona became more dense during the SIMS with an increasing optical depth. Beside the evolution of the Compton corona, the accretion disc was also observed to evolve. The inner disc radius ($R_{\rm in}$) was observed to decrease in the SIMS to $R_{\rm in} \sim 46$~km from $R_{\rm in} \sim 52-59$~km in the HIMS. This indicated that the disc was moving towards to the BH as the outburst progressed.

The QPO frequencies were observed to be correlated with the count rate in the non-dip phases in different energy bands. The correlation of QPO frequency and count rate has been observed in other LMXB also \citep[e.g.,][]{Reig2000,Bogensberger2020}. As the outburst progressed, the flux (count rate) increased and simultaneously, the Compton corona contracts, resulting a higher QPO frequency \citep{CMD2015,AJ2016}. Beside this, we did not observe the QPO frequency-count rate correlation for  dip phases. This could be due to the absorption during the dip phases. The contracting corona also explains the observed anti-correlation between the QPO frequency and HR. The contracting corona would produce less Comptonized emission, compared to the soft-photons, which led to decreasing HR.

The disc wind is equatorial and generally observed in the thermal dominated state, i.e. in SIMS and HSS \citep{Miller2009,Miller2012,Ponti2012}. During the dip phases, an ionized absorber was observed which could be a buldge region at the outer edge of the disc (see Sec~\ref{sec:abs-dips}), which is not seen during the non-dip phases (N1 -- N5). We observed an ionized absorber during the non-dip segment N6 which is clearly different than the absorber that was observed during the dip phases (different $N_{\rm H}$, $\xi$). This absorber could be associated with the disc wind. \citet{Miller2021} also found evidences for disc wind in the XRT spectra of \source~ in the HSS. {The evidence of the disc wind also observed on the optical spectra, which show P-cygni profile H$\alpha$ and He I6678 lines \citep{Mata2022}.}

During our \astrosat~ observation, we found a broad iron emission line ($\sigma \sim 1.2-1.3$~keV). It is possible that more than one lines are blended forming a broad line. As there is evidence of ionized disk wind, ionized Fe lines (Fe H~II line at 6.7~keV or Fe He II line at 6.96~keV) are expected to be observed in the spectrum. However, the spectral resolution of LAXPC did not allow us to resolve them.

\section{Summary}
\label{sec:summary}
We studied a recently discovered black hole candidate MAXI~J1803--298 during its 2021 outburst using the data obtained from \astrosat. We studied the source using the combined data of SXT and LAXPC in the $0.7-80$~keV range. Our key findings are given below.

\begin{enumerate}
\item \astrosat~ observed the presence of periodic dips in the $3-80$~KeV LAXPC light curve with a periodicity of $7.02 \pm 0.18$~hr. The absorption dips are proposed to be caused by the obscured materials of the buldge (thickened material) of the outer accretion disc.
\item We estimate the mass function of the binary as  $f(M) = 2.1-7.2~M_{\odot}$. From this, we estimated the mass of the BH which lies in the range of $M_{\rm BH} \sim 3.5-12.5~M_{\odot}$.
\item The dip and non-dip spectra are fitted with absorbed thermal and Comptonized components. The spectra during the dips are required an addition ionized absorption component. The ionized absorption component is characterized with a column density of $N_{\rm H,2} \sim 28\times 10^{22}$ \pcm~ and ionization parameter, $\xi \sim 10^{3.7}$ \ecps.
\item \astrosat~ observed the source during the state transition. The source was in the HIMS at initial phase of the  observation. The source entered to the SIMS towards the end of our observation. 
\item We find a sharp type-C QPO in every dip and non-dip segments in our observation. We also find that the QPO frequency evolved during our observation duration. 
\item The spectral parameters are also found to evolve during the observation.
\item We find the evidence of evolving Compton corona and accretion disc during our observation. 
\item Evidence of disc wind was observed in the spectra from the SIMS.
\end{enumerate}

\section*{Acknowledgements}
We thank the anonymous reviewer for his/her suggestions and comments that helped us to improve the quality of this manuscript. Work at Physical Research Laboratory, Ahmedabad, is funded by the Department of Space, Government of India. AJ acknowledge the support of the grant from the Ministry of Science and Technology of Taiwan with the grand number MOST 110-2811-M-007-500 and  MOST 111-2811-M-007-002. This research made use of the data obtained through ToO phase of AstroSat observations. The authors thank the SXT-POC of TIFR and the LAXPC team of IUCAA and TIFR for providing the data extraction software for the respective instruments.

\section*{DATA AVAILABILITY}
We used the data of \astrosat~ observatories for this work. 
\bibliographystyle{mnras}
\bibliography{ref-egbhxrb}


\appendix
\section{Table-A}

\begin{table*}
\caption{Energy-Dependent Timing Properties}
\label{tab:en-timing}
\centering
\begin{tabular}{ccccccccccc}
\hline
ID & Energy & Mean Count & QPO$^{\rm f}$ & Q-factor$^{\rm f}$ & rms$^{\rm f}$ & QPO$^{\rm h}$ & Q-factor$^{\rm h}$ & rms$^{\rm h}$ & BN rms \\
& (keV) & (Counts s$^{-1}$) & (Hz) & & (\%) & (Hz) &  & (\%) & (\%) \\
\hline
N1&$3-6  $&$ 445\pm 5$&$ 5.99\pm  0.02$&$ 8.87 \pm 0.09  $&$ 8.13\pm  0.07$&$ 2.94\pm0.06$&$ 2.26\pm 0.10 $&$3.55\pm  0.06$&$ 13.08\pm  1.09$\\ 
&$6-12 $&$ 177\pm 3$&$ 6.01\pm  0.03$&$ 8.39 \pm 0.11  $&$ 8.37\pm  0.08$&$ 2.90\pm0.09$&$ 2.05\pm 0.11 $&$4.46\pm  0.06$&$ 14.21\pm  1.14$\\
&$12-20$&$  64\pm 2$&$ 6.03\pm  0.12$&$ 8.87 \pm 0.05  $&$ 6.41\pm  0.04$&$ 3.23\pm1.12$&$ 3.76\pm 0.30 $&$1.77\pm  0.90$&$ 8.54\pm  1.22$\\
&$20-30$&$  57\pm 2$&$ 6.00\pm  0.16$&$ 6.17 \pm 0.77  $&$ 4.59\pm  0.19$&$ -$&$ - $&$-$&$ 3.10\pm  0.35$\\
&$30-50$&$  51\pm 2$&$ -$&$ -  $&$ -$&$ -$&$ - $&$-$&$ 3.50\pm  0.42$\\
&$50-80$&$  42\pm 2$&$-$&$ -  $&$ -$&$ -$&$ - $&$-$&$ 2.33\pm  0.61$\\
\hline
D1&$3-6  $&$ 402\pm 5$&$ 5.90\pm  0.02$&$ 11.42\pm 0.07  $&$ 7.28\pm  0.50$&$ 2.98\pm0.12$&$ 2.68\pm 0.12 $&$2.68\pm  0.22$&$ 12.53\pm  1.01$\\ 
&$6-12 $&$ 170\pm 3$&$ 5.92\pm  0.03$&$  8.45\pm 0.94  $&$ 7.63\pm  0.47$&$ 3.08\pm0.14$&$ 3.35\pm 0.59 $&$3.48\pm  0.22$&$ 13.67\pm  1.15$\\
&$12-20$&$  61\pm 2$&$ 5.97\pm  0.11$&$  8.42\pm 0.33  $&$ 2.59\pm  0.28$&$ -$&$ - $&$-$&$ 5.25\pm  0.45$\\
&$20-30$&$  60\pm 2$&$ -$&$  -  $&$ -$&$ -$&$ - $&$-$&$ 4.33\pm  0.29$\\
&$30-50$&$  54\pm 2$&$ -$&$  - $&$ -$&$ -$&$ - $&$-$&$ 4.76\pm  0.39$\\
&$50-80$&$  42\pm 2$&$ -$&$  -  $&$ -$&$ -$&$ - $&$-$&$ 1.12\pm  0.18$\\
\hline
N2&$3-6  $&$ 438\pm 5$&$ 5.74\pm  0.02$&$  7.40\pm 0.12  $&$ 8.96\pm  0.12$&$ 2.80\pm0.07$&$ 2.34\pm 0.11 $&$2.72\pm  0.07$&$ 12.98\pm  0.95$\\ 
&$6-12 $&$ 179\pm 3$&$ 5.73\pm  0.02$&$  6.76\pm 0.15  $&$ 9.59\pm  0.11$&$ 2.81\pm0.07$&$ 2.19\pm 0.13 $&$3.74\pm  0.12$&$ 15.67\pm  1.08$\\
&$12-20$&$  66\pm 2$&$ 5.77\pm  0.04$&$  9.07\pm 0.11  $&$ 7.06\pm  0.53$&$ 3.02\pm0.11$&$ 2.34\pm 0.20 $&$1.82\pm  0.32$&$ 11.97\pm  1.09$\\
&$20-30$&$  61\pm 2$&$ 5.91\pm  0.10$&$ 10.53\pm 0.27  $&$ 4.89\pm  0.21$&$ -$&$ - $&$-$&$ 9.34\pm  0.83$\\
&$30-50$&$  55\pm 2$&$- $&$ -   $&$ -$&$ -$&$ - $&$-$&$ 5.44\pm  0.19$\\
&$50-80$&$  44\pm 2$&$- $&$  -  $&$ -$&$ -$&$ - $&$-$&$ 4.41\pm  0.27$\\
\hline
D2&$3-6  $&$ 375\pm 4$&$ 5.91\pm  0.03$&$  9.89\pm 0.03  $&$ 8.79\pm  0.04$&$ 2.87\pm0.18$&$ 2.42\pm 0.10 $&$2.37\pm  0.18$&$ 12.98\pm  1.14$\\ 
&$6-12 $&$ 166\pm 3$&$ 5.92\pm  0.03$&$ 10.19\pm 0.12  $&$ 9.63\pm  0.09$&$ -$&$ - $&$-$&$ 10.21\pm  1.18$\\
&$12-20$&$  57\pm 2$&$ 5.71\pm  0.15$&$ 11.49\pm 0.33  $&$ 4.25\pm  0.26$&$ -$&$ - $&$-$&$ 5.23\pm  0.51$\\
&$20-30$&$  55\pm 1$&$ -$&$  - $&$ -$&$ -$&$ -$&$-$&$ 3.53\pm  0.41$\\
&$30-50$&$  50\pm 2$&$ -$&$  -  $&$ -$&$ -$&$ - $&$-$&$ 3.56\pm  0.61$\\
&$50-80$&$  39\pm 2$&$ -$&$  -  $&$ -$&$ -$&$ - $&$-$&$ 2.69\pm  0.14$\\
\hline
N3&$3-6  $&$ 434\pm 5$&$ 5.64\pm  0.01$&$  6.24\pm 0.18  $&$ 6.79\pm  0.16$&$ 2.75\pm0.04$&$ 2.66\pm 0.14 $&$2.34\pm  0.08$&$ 10.94\pm  1.08$\\
&$6-12 $&$ 177\pm 3$&$ 5.68\pm  0.02$&$  5.86\pm 0.22  $&$ 7.16\pm  0.14$&$ 2.64\pm0.07$&$ 1.99\pm 0.15 $&$3.54\pm  0.10$&$ 13.12\pm  1.11$\\
&$12-20$&$  68\pm 2$&$ 5.58\pm  0.05$&$  5.43\pm 0.26  $&$ 3.87\pm  0.15$&$ 3.23\pm0.21$&$ 4.57\pm 0.29 $&$1.97\pm  0.31$&$ 8.05\pm  0.95$\\
&$20-30$&$  57\pm 1$&$ 5.55\pm  0.11$&$  3.76\pm 0.32  $&$ 3.37\pm  0.21$&$ -$&$ - $&$-$&$ 6.63\pm  0.24$\\
&$30-50$&$  50\pm 2$&$ -$&$  -  $&$ -$&$ -$&$ - $&$-$&$ 3.62\pm  0.39$\\
&$50-80$&$  39\pm 2$&$ -$&$  -  $&$ -$&$ -$&$ - $&$-$&$ 2.56\pm  0.12$\\
\hline
D3&$3-6  $&$ 355\pm 4$&$ 5.39\pm  0.03$&$ 10.12\pm 0.07  $&$ 6.83\pm  0.05$&$ 2.33\pm0.18$&$ 1.73\pm 0.17 $&$2.76\pm  0.25$&$ 11.69\pm  1.09$\\
&$6-12 $&$ 165\pm 3$&$ 5.40\pm  0.04$&$  8.12\pm 0.11  $&$ 7.54\pm  0.08$&$ 2.81\pm0.07$&$ 7.34\pm 0.17 $&$3.06\pm  0.19$&$11.76\pm  1.20$\\
&$12-20$&$  61\pm 2$&$ 5.23\pm  0.12$&$  9.58\pm 0.31  $&$ 4.03\pm  0.23$&$ -$&$ - $&$-$&$ 6.23\pm  0.23$\\
&$20-30$&$  55\pm 1$&$ -$&$  -  $&$ -$&$ -$&$ - $&$-$&$ 3.12\pm  0.21$\\
&$30-50$&$  49\pm 2$&$ -$&$  -  $&$ -$&$ -$&$ - $&$-$&$ 2.33\pm  0.20$\\
&$50-80$&$  36\pm 2$&$ -$&$  -  $&$ -$&$ -$&$ - $&$-$&$ 1.65\pm  0.13$\\
\hline

N4&$3-6  $&$ 417\pm 5$&$ 5.30\pm  0.02$&$  7.06\pm 0.05  $&$ 5.43\pm  0.03$&$ 2.46\pm0.06$&$ 2.45\pm 0.08 $&$2.27\pm  0.10$&$ 10.64\pm  1.08$\\ 
&$6-12 $&$ 175\pm 3$&$ 5.29\pm  0.02$&$  6.83\pm 0.06  $&$ 7.71\pm  0.04$&$ 2.51\pm0.13$&$ 1.93\pm 0.14 $&$3.10\pm  0.22$&$ 13.59\pm  1.22$\\
&$12-20$&$  69\pm 2$&$ 5.35\pm  0.12$&$  6.40\pm 0.24  $&$ 6.19\pm  0.20$&$ -$&$ - $&$-$&$ 7.57\pm  0.69$\\
&$20-30$&$  58\pm 1$&$ 5.33\pm  0.13$&$  5.14\pm 0.21  $&$ 3.13\pm  0.34$&$ -$&$ - $&$-$&$ 6.00\pm  0.46$\\
&$30-50$&$  50\pm 2$&$ -$&$  -  $&$ -$&$ -$&$ - $&$-$&$ 1.14\pm  0.21$\\
&$50-80$&$  39\pm 2$&$ -$&$  -  $&$ -$&$ -$&$ - $&$-$&$ 1.80\pm  0.18$\\
\hline
D4&$3-6  $&$ 372\pm 4$&$ 6.24\pm  0.03$&$  8.43\pm 0.12  $&$ 5.51\pm  0.08$&$ 3.05\pm0.14$&$ 2.74\pm 0.21 $&$2.80\pm  0.23$&$ 10.90\pm  1.16$\\
&$6-12 $&$ 161\pm 3$&$ 6.31\pm  0.06$&$  8.37\pm 0.17  $&$ 6.53\pm  0.11$&$ 2.97\pm0.18$&$ 3.84\pm 0.29 $&$3.25\pm  0.34$&$ 11.14\pm  1.18$\\
&$12-20$&$  59\pm 2$&$ 6.30\pm  0.22$&$ 12.62\pm 0.75  $&$ 2.26\pm  0.51$&$ -$&$ - $&$-$&$ 5.41\pm  0.43$&\\
&$20-30$&$  57\pm 1$&$ -$&$  -  $&$ -$&$ -$&$ - $&$-$&$ 3.34\pm  0.46$\\
&$30-50$&$  49\pm 2$&$ -$&$  -  $&$ -$&$ -$&$ - $&$-$&$ 3.17\pm  0.55$\\
&$50-80$&$  35\pm 2$&$ -$&$  -  $&$ -$&$ -$&$ - $&$-$&$ 2.56\pm  0.16$\\
\hline
N5&$3-6  $&$ 492\pm 5$&$ 6.77\pm  0.08$&$  6.10\pm 0.21  $&$ 5.98\pm  0.13$&$ 3.16\pm0.24$&$ 2.00\pm 0.31 $&$3.30\pm  0.25$&$ 12.96\pm  1.10$\\
&$6-12 $&$ 188\pm 3$&$ 6.78\pm  0.10$&$  4.81\pm 0.24  $&$ 7.52\pm  0.14$&$ 3.71\pm0.22$&$ 3.07\pm 0.35 $&$4.34\pm  0.32$&$ 12.56\pm  1.15$\\
&$12-20$&$  71\pm 2$&$ 7.05\pm  0.22$&$  5.12\pm 0.87  $&$ 5.50\pm  0.53$&$ -$&$ - $&$-$&$ 7.89\pm  0.78$\\
&$20-30$&$  61\pm 1$&$ -$&$  -  $&$ -$&$ -$&$ - $&$-$&$ 5.10\pm  0.15$&\\
&$30-50$&$  55\pm 2$&$ -$&$  -  $&$ -$&$ -$&$ - $&$-$&$ 4.39\pm  0.32$&\\
&$50-80$&$  41\pm 2$&$ -$&$  -  $&$ -$&$ -$&$ - $&$-$&$ 2.23\pm  0.18$&\\
\hline
\end{tabular}
\end{table*}

\begin{table*}
\contcaption{Energy-Dependent Timing Properties}
\centering
\begin{tabular}{ccccccccccc}
\hline
ID & Energy & Mean Count & QPO$^{\rm f}$ & Q-factor$^{\rm f}$ & rms$^{\rm f}$ & QPO$^{\rm h}$ & Q-factor$^{\rm h}$ & rms$^{\rm h}$ & BN rms \\
& (keV) & (Counts s$^{-1}$) & (Hz) & & (\%) & (Hz) &  & (\%) & (\%) \\
\hline
N6&$3-6  $&$ 558\pm 5$&$ 7.54\pm  0.07$&$  9.46\pm 0.22  $&$ 8.10\pm  0.14$&$ -$&$ - $&$-$&$ 8.50\pm  0.91$\\ 
&$6-12 $&$ 203\pm 3$&$ 7.66\pm  0.26$&$  3.92\pm 0.49  $&$ 9.65\pm  0.27$&$ -$&$ - $&$-$&$ 9.92\pm  1.17$\\
&$12-20$&$  69\pm 2$&$ -$&$  -  $&$ -$&$ -$&$ - $&$-$&$ 9.98\pm  0.42$\\
&$20-30$&$  60\pm 2$&$ -$&$  -  $&$ -$&$ -$&$ - $&$-$&$ 6.01\pm  0.32$\\
&$30-50$&$  56\pm 2$&$ -$&$  -  $&$ -$&$ -$&$ - $&$-$&$ 4.69\pm  0.48$\\
&$50-80$&$  46\pm 2$&$ -$&$  -  $&$ -$&$ -$&$ - $&$-$&$ 2.13\pm  0.15$\\
\hline
\end{tabular}
\leftline{The parameters corresponding to the fundamental component are indicated with superscript `f' and those corresponding to the sub-harmonic component with `h'.}
\end{table*}

\begin{table*}
\caption{Spectral Analysis Result}
\centering
\begin{tabular}{lcccccccccccccccc}
\hline
ID & Date & $N_{\rm H,1}$ & $N_{\rm H,2}$ & $\log~\xi$ & CF & $T_{\rm in}$ & $N_{\rm dbb}$ & $\Gamma$ & $kT_{\rm e}$  \\
 & (Day*) & ($10^{22}$ \pcm) & ($10^{22}$ \pcm) & $\log$~(\ecps) & & (keV) & & & (keV) \\
(1) & (2) & (3) & (4) & (5) & (6) & (7) & (8) & (9) & (10) \\
\\
 & & $\tau$ &$N_{\rm nth}$ & $\sigma$ & EW & LN & $F_{\rm tot}$ & $f_{\rm disc}$ & $\chi^2$/dof  \\
 & & &($10^{-2}$ \phc) & (keV) & (keV) & ($10^{-2}$ \phc) & ($10^{-9}$ \ecs) &  & \\
 & & (11) & (12) & (13) & (14) & (15) & (16) & (17) & (18)\\
\hline
N1& 0.58& $ 0.41^{+0.02}_{-0.01}$ & $       -              $ & $      -               $ & $ -                    $ & $0.88^{+0.02}_{-0.02}$ & $ 1236^{+95}_{-78} $ & $ 2.04^{+0.04}_{-0.03}$ & $ 37^{+22}_{-10}$ \\
\\
& & $ 1.98^{+0.54}_{-0.64}$& $0.49^{+0.04}_{-0.04}$ & $ 1.24^{+0.21}_{-0.16}$ & $ 0.54^{+0.01}_{-0.02}$ & $ 3.58^{+0.89}_{-0.64}$ & $  9.18^{+0.08}_{-0.09}$ & $ 0.43^{+0.02}_{-0.02}$ &  659/644 \\
\hline
D1& 0.68& $ 0.40^{+0.02}_{-0.02}$ & $ 28.78^{+8.76}_{-6.21}$ & $  3.78^{+0.63}_{-0.34}$ & $ 0.43^{+0.02}_{-0.03} $ & $0.86^{+0.01}_{-0.01}$ & $ 1488^{+66}_{-116}$ & $ 2.05^{+0.04}_{-0.05}$ & $ 35^{+15}_{-22}$ \\
\\
& & $ 2.04^{+1.09}_{-0.49}$& $0.33^{+0.03}_{-0.04}$ & $ 1.26^{+0.15}_{-0.19}$ & $ 0.62^{+0.02}_{-0.01}$ & $ 5.07^{+0.94}_{-1.45}$ & $  8.71^{+0.07}_{-0.10}$ & $ 0.42^{+0.02}_{-0.03}$ &  683/654\\
\hline
N2& 0.82& $ 0.45^{+0.01}_{-0.03}$ & $  -                   $ & $ -                    $ & $  -                   $ & $0.89^{+0.02}_{-0.02}$ & $ 1272^{+92}_{-45} $ & $ 2.06^{+0.04}_{-0.02}$ & $ 35^{+17}_{-24}$ \\
\\
& & $ 2.01^{+1.44}_{-0.45}$& $0.38^{+0.02}_{-0.05}$ & $ 1.18^{+0.12}_{-0.25}$ & $ 0.71^{+0.01}_{-0.01}$ & $ 4.27^{+0.96}_{-0.68}$ & $  9.07^{+0.11}_{-0.13}$ & $ 0.42^{+0.02}_{-0.03}$ &  672/633\\
\hline
D2& 0.99& $ 0.39^{+0.03}_{-0.02}$ & $ 27.99^{+6.21}_{-8.12}$ & $  3.73^{+0.58}_{-0.43}$ & $ 0.45^{+0.03}_{-0.03} $ & $0.90^{+0.01}_{-0.02}$ & $ 1305^{+53}_{-71} $ & $ 2.01^{+0.03}_{-0.03}$ & $ 36^{+26}_{-15}$ \\
\\
& & $ 2.06^{+0.97}_{-0.71}$& $0.31^{+0.03}_{-0.02}$ & $ 1.21^{+0.16}_{-0.21}$ & $ 0.48^{+0.01}_{-0.02}$ & $ 3.77^{+1.06}_{-0.54}$ & $  7.75^{+0.09}_{-0.12}$ & $ 0.42^{+0.01}_{-0.03}$ &  648/630\\
\hline
N3& 1.14& $ 0.45^{+0.02}_{-0.04}$ & $  -                   $ & $    -                 $ & $ -                    $ & $0.91^{+0.04}_{-0.02}$ & $ 1439^{+103}_{-83}$ & $ 2.03^{+0.03}_{-0.04}$ & $ 36^{+11}_{-16}$ \\
\\
&& $ 2.04^{+1.08}_{-0.41}$ & $0.39^{+0.04}_{-0.03}$ & $ 1.23^{+0.13}_{-0.18}$ & $ 0.60^{+0.02}_{-0.03}$ & $ 4.76^{+0.91}_{-1.15} $ & $  9.07^{+0.11}_{-0.08}$ & $ 0.43^{+0.02}_{-0.02}$ &  689/659\\
\hline
D3& 1.29& $ 0.40^{+0.02}_{-0.03}$ & $ 28.94^{+5.23}_{-10.73}$ & $ 3.78^{+0.64}_{-0.49}$ & $ 0.55^{+0.02}_{-0.04} $ & $0.89^{+0.02}_{-0.01}$ & $ 1378^{+98}_{-77} $ & $ 2.03^{+0.03}_{-0.05}$ & $ 36^{+14}_{-19}$ \\
\\
& & $ 2.03^{+1.25}_{-0.59}$& $0.27^{+0.04}_{-0.05}$ & $ 1.20^{+0.15}_{-0.11}$ & $ 0.57^{+0.02}_{-0.02}$ & $ 4.08^{+0.75}_{-0.92}$ & $  7.32^{+0.12}_{-0.09}$ & $ 0.41^{+0.03}_{-0.02}$ &  714/701\\
\hline
N4& 1.43& $ 0.39^{+0.03}_{-0.01}$ & $  -                   $ & $        -             $ & $     -                $ & $0.87^{+0.01}_{-0.03}$ & $ 1312^{+112}_{91} $ & $ 2.03^{+0.04}_{-0.02}$ & $ 36^{+19}_{-12}$ \\
\\
& & $ 2.04^{+0.70}_{-0.60}$& $0.46^{+0.03}_{-0.03}$ & $ 1.20^{+0.15}_{-0.15}$ & $ 0.65^{+0.01}_{-0.01}$ & $ 4.67^{+0.84}_{-1.17}$ & $  9.19^{+0.15}_{-0.12}$ & $ 0.42^{+0.01}_{-0.02}$ &  701/642\\
\hline
D4& 1.58& $ 0.42^{+0.02}_{-0.02}$ & $ 28.07^{+7.22}_{-9.31}$ & $  3.74^{+0.54}_{-0.59}$ & $ 0.45^{+0.03}_{-0.04} $ & $0.88^{+0.02}_{-0.02}$ & $ 1213^{+78}_{-109}$ & $ 2.04^{+0.02}_{-0.03}$ & $ 36^{+21}_{-17}$ \\
\\
&& $ 2.00^{+1.20}_{-0.58}$ & $0.33^{+0.02}_{-0.03}$ & $ 1.08^{+0.17}_{-0.12}$ & $ 0.61^{+0.01}_{-0.01}$ & $ 4.41^{+0.78}_{-0.96}$ & $  8.15^{+0.15}_{-0.09}$ & $ 0.42^{+0.02}_{-0.02}$ &  655/627\\
\hline
N5& 1.66& $ 0.40^{+0.02}_{-0.02}$ & $  -                   $ & $        -             $ & $      -               $ & $0.97^{+0.02}_{-0.04}$ & $ 1028^{+65}_{-88} $ & $ 2.08^{+0.03}_{-0.04}$ & $ 31^{+21}_{-10}$ \\
\\
&& $ 2.19^{+0.72}_{-0.73}$ & $0.43^{+0.05}_{-0.05}$ & $ 1.27^{+0.08}_{-0.20}$ & $ 0.51^{+0.01}_{-0.02}$ & $ 3.63^{+0.54}_{-0.82}$ & $ 10.62^{+0.10}_{-0.14}$ & $ 0.45^{+0.03}_{-0.02}$ &  703/665\\
\hline
N6& 1.72& $ 0.46^{+0.01}_{-0.03}$ & $  7.61^{+2.23}_{-1.23}$ & $  3.67^{+0.71}_{-0.33}$ & $ 0.53^{+0.02}_{-0.05} $ & $1.03^{+0.03}_{-0.02}$ & $  951^{+81}_{-63} $ & $ 2.12^{+0.02}_{-0.03}$ & $ 29^{+18}_{-11}$ \\
\\
& & $ 2.21^{+0.92}_{-0.68}$& $0.46^{+0.04}_{-0.03}$ & $ 1.16^{+0.21}_{-0.15}$ & $ 0.48^{+0.02}_{-0.01}$ & $ 3.86^{+1.05}_{-0.77}$ & $ 11.75^{+0.11}_{-0.16}$ & $ 0.48^{+0.02}_{-0.03}$ &  702/639\\
\hline
\end{tabular}
\leftline{* Day 0 = MJD 59345. Errors are quoted at 1 $\sigma$.}
\label{tab:spec}
\end{table*}



\bsp	
\label{lastpage}
\end{document}